\documentclass[a4paper,11pt]{article}
\pdfoutput=1

\usepackage[a4paper,left=2.73cm,right=2.7cm,top=3cm,bottom=3.5cm]{geometry}
\usepackage[T1]{fontenc} 
\usepackage{graphicx}
\usepackage{epsfig}
\usepackage{amssymb}
\usepackage{relsize}
\usepackage{amsfonts}
\usepackage{dsfont}
\usepackage{amsmath,euscript,array,mathrsfs,nccmath}
\usepackage{multirow}
\usepackage{bbold,bm,bbm}
\usepackage{epsf}
\usepackage{slashed}
\usepackage{comment}
\usepackage{colortbl}
\usepackage{lmodern}
\usepackage[utf8]{inputenc}
\usepackage[noadjust]{cite}
\usepackage{xcolor}
\usepackage{jheppub}
\usepackage{lmodern}
\usepackage{caption,subcaption,wrapfig}
\usepackage[colorlinks=true,linktocpage=true,linkcolor=blue,citecolor=blue]{hyperref}
\usepackage{tikz}
\usetikzlibrary{decorations.markings,decorations.pathmorphing}

\newcommand{\wz}{w_0}

\newcommand{\dd}{\text{d}}
\newcommand{\parent}[1]{\left(#1\right)}

\newcommand{\nrho}{\varrho}
\newcommand{\fs}{\Lambda}
\newcommand{\fv}{v}
\newcommand{\st}{\sigma}
\newcommand{\ratioT}{\mathcal{T}}

\newcommand{\zt}{\tilde{z}}

\newcommand{\OO}{\mathcal{O}}
\newcommand{\Qtt}{Q_{1}}
\newcommand{\Qxx}{Q_{2}}
\newcommand{\Qyy}{Q_{3}}
\newcommand{\Qzz}{Q_{4}}
\newcommand{\Qzx}{Q_{5}}
\newcommand{\Qphi}{Q_0}
\newcommand{\qtt}{q_{tt}}
\newcommand{\qxx}{q_{\rho\rho}}
\newcommand{\ftwo}{q_\phi}
\newcommand{\qfive}{q_{\rho z}}
\newcommand{\spacelist}{1mm}
\newcommand{\PL}{p_{||}}
\newcommand{\PT}{p_{\perp}}
\newcommand{\VEV}{\langle\OO\rangle}
\newcommand{\source}{\phi_s}
\newcommand{\mn}{{\mu\nu}}
\newcommand{\Rb}{\bar{R}}
\newcommand{\Rh}{{\tilde{R}}}
\newcommand{\nb}{\bar{\nabla}}
\newcommand{\eUV}{\epsilon_\text{\tiny UV}}
\newcommand{\homphi}{\psi}
\newcommand{\radius}{R}
\newcommand{\widthGaussian}{b}
\newcommand{\kH}{\kappa_{\text{\tiny H}}}
\newcommand{\field}{\varphi}
\newcommand{\Tc}{T_{\text{c}}}

\newcommand{\zH}{z_{\text{\tiny H}}}
\newcommand{\gH}{g_{\text{\tiny H}}}
\newcommand{\fH}{f_{\text{\tiny H}}}
\newcommand{\rhot}{\tilde\rho}

\newcommand{\Veff}{V_{\text{eff}}}
\newcommand{\ho}{0}
\newcommand{\lw}{\ell}
\newcommand{\eqn}[1]{(\ref{#1})}
\newcommand{\sqb}{\bar\square}
\newcommand{\cnta}{\mathsf{a}}
\newcommand{\cntd}{\mathsf{b}}
\newcommand{\cntt}{\mathsf{c}}
\newcommand{\be}{\beta}

\begin{document}
	
	\definecolor{color1}{rgb}{0.117647, 0.564706, 1.}
	\definecolor{color2}{rgb}{0.411765, 0.466667,0.666667}
	\definecolor{color3}{rgb}{0.705882, 0.368627, 0.333333}
	\definecolor{color4}{rgb}{1,0.270588,0}
	
	\begin{titlepage}
		
		\thispagestyle{empty}
		
		\mbox{}
		
		\vspace{20pt}  
		
		\begin{center}

			\begin{center}
				{\Huge \textbf{Microscopic Description}}\\ \vspace{5mm} 
				{\Huge \textbf{of }}\\ \vspace{5mm} 
				{\Huge\textbf{Critical Bubbles}}
			\end{center}

			\vspace{50pt}
			{
				\bf{Carlos Hoyos,}$^{1, 2}$
				\bf{David Mateos,}$^{3,4,5}$\\
				\vspace{2mm}
				\bf{Wilke van der Schee,}$^{6,7,8}$ \bf{Javier G. Subils}$^{6}$
			}

			\vspace{20pt}

			\normalsize 
			{\small  $^{1}$ \textit{Departamento de F\'{i}sica, Universidad de Oviedo, \\ Calle Leopoldo Calvo Sotelo 18, ES-33007, Oviedo, Spain.}}\\
			{ $^{2}$ \textit{Instituto Universitario de Ciencias y Tecnolog\'{\i}as Espaciales de Asturias (ICTEA), \\ Calle de la Independencia 13, ES-33004, Oviedo, Spain.}}
			\\
			{\small $^{3}$\textit{
					Departament de Física Quàntica i Astrofísica \& Institut de Ciències del Cosmos (ICC), Universitat de Barcelona, 
					Martí i Franquès 1, ES-08028, Barcelona, Spain.}}\\
			{\small $^{5}$\textit{
					Institució Catalana de Recerca i Estudis Avançats (ICREA),\\ Passeig Lluís Companys 23, ES-08010, Barcelona, Spain.}}\\
			{\small $^{6}$\textit{
					Institute for Theoretical Physics, Utrecht University, 3584 CC Utrecht, The Netherlands.}}\\
			{\small $^{7}$\textit{
					Theoretical Physics Department, CERN CH-1211 Genève 23, Switzerland.}}\\
			{\small $^{8}$\textit{Nikhef, Science Park 105, 1098 XG Amsterdam, The Netherlands.}}\\
			\vspace{15pt}
			\vspace{30pt}
			\textbf{Abstract}
		\end{center} 
		{\small
			First-order phase transitions occur through  the nucleation of critical bubbles of the stable phase within the metastable phase. Using holography, we present a fully microscopic description of these bubbles in a strongly coupled, four-dimensional gauge theory at finite temperature. In the gravitational dual, these bubbles correspond to static, inhomogeneous and unstable black-brane solutions with a localized deformation on the horizon. We construct these solutions across the entire metastable branch and compute the surface tension and the nucleation rate. We then compare these microscopic results with those obtained from a two-derivative effective action for the order parameter in two different scenarios. When the effective action is derived from the microscopic theory via holography, we find remarkable agreement. However, when the effective action is constrained only by the equation of state and dimensional analysis, significant discrepancies emerge. These discrepancies can be resolved if an additional constraint related to the surface tension is imposed.}
		
	\end{titlepage}	
	
	\newpage
	\vspace{5mm}
	
	\setcounter{page}{2}
	
	\tableofcontents
	
	\section{Introduction}
	First-order phase transitions (FOPT) are a ubiquitous phenomenon across physics. They play a central role in a wide range of systems, from condensed matter to high-energy contexts. In cosmology, for instance, FOPT may occur during the early universe in many scenarios that extend the Standard Model \cite{Linde:1978px,Kibble:1980mv,Enqvist:1991xw,Mazumdar:2018dfl,Hindmarsh:2020hop} (see \cite{Caprini:2019egz} for a recent review). This has potential consequences for baryogenesis \cite{Kuzmin:1985mm,Shaposhnikov:1986jp,McLerran:1991zh,Farrar:1993sp,Rubakov:1996vz}, gravitational wave production \cite{Witten:1984rs,Hogan:1986qda,Kosowsky:1991ua,Kosowsky:1992rz,Kamionkowski:1993fg,Caprini:2009yp,Caprini:2015zlo,Hindmarsh:2013xza,Hindmarsh:2015qta,Athron:2023xlk}, and cosmic defect formation \cite{Kibble:1976sj,Kibble:1980mv,Zurek:1985qw,Rajantie:2003vv,Kawasaki:2014sqa,Vachaspati:2006zz}. In astrophysics, they are relevant to the physics of dense QCD matter in core collapse supernovae 
	\cite{Gentile:1993ma,Hong:2001gt,Fischer:2010wp,Sagert2011,Fischer:2011zj,Nishimura:2011yb} or in neutron stars (mergers) 
	\cite{Glendenning:1992vb,Alford:2004pf,Bauswein:2018bma,Most:2018eaw,Most:2018hfd,Ecker:2019xrw,Prakash:2021wpz, Weih:2019xvw,Tootle:2022pvd,Casalderrey-Solana:2022rrn}.
	In all these settings, the dynamics of the transition is governed by a relativistic Quantum Field Theory (QFT) at finite temperature.
	
	A key feature of FOPT is the nucleation of critical bubbles of the stable phase within a metastable background. These  bubbles are intrinsically nonperturbative objects whose properties govern important aspects of the transition dynamics and its observable consequences. The free energy of the critical bubbles determines the nucleation rate, which in turn influences the duration and completion of the phase transition \cite{Linde:1981zj, Coleman:1977py}. The bubble wall profile, describing the spatial variation of the order parameter and other fields across the bubble interface, plays a crucial role in scenarios where baryogenesis occurs through a FOPT, by controlling  the necessary CP-violating and out-of-equilibrium effects  \cite{Cohen:1993nk, Morrissey:2012db, Joyce:1994zn,vandeVis:2025efm}. Understanding these microscopic features is therefore essential for a precise modeling of the phenomenology of FOPTs.
	
	In most applications, the critical bubbles are studied using an effective field theory (EFT) description for the order parameter, which is usually assumed to be a single scalar field. While this approach can provide powerful insights, it relies on the validity of a derivative expansion that may not hold if gradients become large, as expected at the bubble wall. In addition, in many cases of interest the effective action cannot be determined from first principles. In some instances, this is because the microscopic theory is unknown, as in Beyond the Standard Model scenarios. In other cases, the microscopic theory is known, but calculating the effective action from first principles is not feasible due to technical challenges, such as the strongly coupled nature of the physics. Under these circumstances, one can still attempt to construct an effective action motivated by a combination of phenomenological considerations and dimensional analysis. To assess the regime of validity of these approximations, a fully microscopic description of the critical bubbles is necessary. This requires a nonperturbative treatment of the underlying QFT, regardless of whether the theory is weakly or strongly coupled. 
	
	In this work, we pursue such a microscopic approach using  holography, which provides a framework for studying large-$N$, strongly coupled gauge theories via a dual gravitational description in higher dimensions \cite{Maldacena:1997re,Witten:1998qj,Gubser:1998bc}. In particular, thermal phase transitions in the boundary theory correspond to transitions between black brane geometries in the bulk.  This approach has been successfully used to assess the validity of hydrodynamics \cite{Heller:2007qt,Chesler:2007an,Chesler:2007sv,Chesler:2008uy,Chesler:2009cy,Chesler:2010bi,Heller:2011ju,Casalderrey-Solana:2013aba}, to simulate the real-time dynamics of a phase transition \cite{Attems:2017ezz,Janik:2017ykj,Attems:2018gou,Attems:2019yqn, Bellantuono:2019wbn, Bea:2021zol,Janik:2021jbq}, to determine the velocity with which supercritical bubbles expand \cite{Bea:2021zsu,Bigazzi:2021ucw,Bea:2022mfb,Bea:2024bls}, to compute the corresponding gravitational wave spectrum \cite{Ares:2020lbt,Bea:2021zol}, and to find the effective action for the order parameter \cite{Ares:2021nap,Ares:2021ntv}. 
	
\begin{figure}[t]
		\centering
		\includegraphics[width=0.70\linewidth]{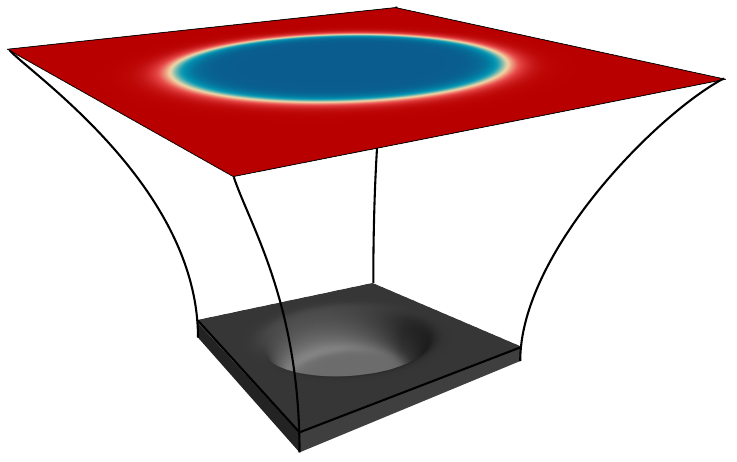}
		\caption{\small A critical bubble (top part of the figure) is dual to an inhomogeneous static black brane geometry (bottom part).}
		\label{fig:bubble_picture}
	\end{figure}

	Ref.~\cite{Bea:2022mfb} provided an approximate microscopic description of critical bubbles by fine-tuning the initial conditions in a time-dependent set-up. More recently, the problem has been investigated in the probe approximation, in which only a small subset of degrees of freedom undergo the FOPT, with the goal of testing the  effective approach \cite{Henriksson:2025vci}, finding good agreement. In contrast, here we  construct  fully backreacted, exact duals of critical bubbles as static, inhomogeneous, and unstable black-brane solutions featuring localized deformations along the horizon---see Fig.~\ref{fig:bubble_picture}.
	This approach  allows us to compute the properties of the bubbles directly from the microscopic theory. We  then compare these properties to those obtained from a two-derivative effective action for the order parameter  in two different scenarios. In the first one, we use holography to derive the effective action from first principles. In this case we find remarkable agreement with the microscopic results. In the second scenario, we simply model the effective potential as a degree-four polynomial, determined by the equation of state, and fix the coefficient of the kinetic term using dimensional analysis. This approach significantly overestimates the magnitude of this coefficient, leading to substantial discrepancies with the microscopic results. Ultimately, these discrepancies can be traced to a suppression of the surface tension relative to the naive dimensional estimate. By adjusting the kinetic term in the effective action to ensure that the correct surface tension is reproduced, we restore agreement with the microscopic results.
	
	Our results provide new insights into the microscopic structure of phase transitions in strongly coupled systems, offering a controlled framework for testing and extending the applicability of effective models. The strong interactions suggest parallels with FOPT in pure $SU(N)$ Yang-Mills (YM) theories with $N > 2$. While we will comment on these similarities below,  it is important to interpret this comparison with caution, as, unlike in YM theories, the FOPT in our case occurs between two plasma-like, deconfined phases.
	
	In this paper we will focus on so-called $O(3)$-symmetric bubbles, namely, those nucleated mostly through thermal fluctuations, as opposed to $O(4)$-symmetric bubbles nucleated mostly through quantum effects. In our case, it is plausible that the former dominate the nucleation because the metastable branch does not extend to parametrically small temperatures. Since $O(3)$-symmetric bubbles are independent of the Euclidean-time direction, we will find them directly as static Lorentzian solutions, which should be thought of as initial conditions for the post-nucleation Lorentzian-time evolution. 
	
	The paper is organized as follows. In Section~\ref{sec:holo_setup} we introduce the model, present its main properties, and explain how to construct homogeneous and inhomogeneous phases. In Section~\ref{sec:results} we present our results for the critical bubbles at different temperatures. In Section~\ref{sec:effective} we investigate if it is possible to obtain similar results from an effective approach. We conclude and discuss future directions in Section~\ref{sec:conclusions}. Technical details of our computations an numerical implementation are relegated to Appendices~\ref{app:implementation},~\ref{app:holoren},~\ref{app:free energy},~\ref{app:effact} and~\ref{app:greenfunc}.
	
	\section{Holographic set-up}
	\label{sec:holo_setup}
	
	\subsection{Model}
	\label{sec:model}
	Our gravitational bulk theory is described by the Einstein-dilaton action 
	\begin{equation}\label{eq:actionGH}
		I_g\,=\,\frac{2}{\kappa_5^2}\int \dd^5x\sqrt{-G}\left(\frac{R}{4}-\frac12 \partial_M\phi\partial^M\phi
		-V(\phi)\right)\,, 
	\end{equation}
	with $G$ the determinant of the five-dimensional spacetime metric $G_{MN}$ and $R$ its Ricci scalar. We assume that the potential $V(\phi)$ can be written in terms of a superpotential $W(\phi)$ through the standard relation 
	\begin{equation}
		\label{eq:pot.from.superpot}
		V(\phi)= -\frac{4}{3}W(\phi)^2 + \frac{1}{2}\left(\frac{\partial W(\phi)}{\partial \phi}\right)^2\,.
	\end{equation}
	This choice does not imply that the theory is supersymmetric but it is simply a convenient way to specify the actual potential. We take the superpotential to lie in the class of models introduced in \cite{Bea:2018whf}:
	\begin{equation}
		W(\phi) = \frac{1}{L}\parent{-\frac{3}{2} - \frac{\phi^2}{2} + \lambda_4 \phi^4 + \lambda_6 \phi^6}\,,
	\end{equation}
	where $L$ is the asymptotic AdS radius,   $\lambda_4=-1/4$ and $\lambda_6=1/10$. In the notation of Refs.~\cite{Bea:2020ees,Bea:2021zsu,Bea:2021zol,Bea:2022mfb}, these choices correspond to $\phi_M=1, \phi_Q=10$, and are motivated by the fact that  they give rise to a simple bulk theory dual to a four-dimensional gauge theory with a first-order phase transition.  More details can be found in Ref.~\cite{Bea:2018whf}. 
	
	The potential has a maximum at the origin, where the system admits as a solution AdS space with radius $L$. This solution is dual to a 3+1 dimensional conformal field theory (CFT). The potential approaches the maximum as
	\begin{equation}\label{eq:expansion_potential}
		V(\phi) = \frac{1}{L^ 2}\parent{-3 -\frac{3}{2}\phi^2-\frac{\phi^4}{3}+O(\phi^6)}\,.
	\end{equation}
	In particular, the mass of $\phi$ around $\phi=0$ is $m^ 2L^ 2 = -3$, which means that solutions with nontrivial profiles for this scalar field describe deformations of the CFT  by a dimension-three scalar operator $\OO$, dual to~$\phi$. In addition, for our choice of parameters $\lambda_4$ and $\lambda_6$, the potential and the superpotential have a minimum at 
	\begin{equation}
		\phi^ 2_{\text{\tiny IR}} = {-\frac{\lambda_4}{3\lambda_6} + \frac{\sqrt{4\lambda_4^2 + 6\lambda_6}}{6\lambda_6}}\,,
	\end{equation}
	where AdS with radius $L_{\text{\tiny IR}} = \sqrt{-3/V(\phi_{\text{\tiny IR}})}$ is again an exact solution, dual to an IR CFT to which the deformed UV theory flows. For different values of the parameters additional extrema can appear, leading to interesting dynamics \cite{Bea:2018whf}.
	
	Finite-temperature states in the gauge theory are dual in the bulk to solutions with a horizon. We will discuss how to find these in the rest of this section. In Section~\ref{sec:homogeneous} we focus on homogeneous phases, described by homogeneous black brane solutions. We know from previous work that some of these will be metastable or unstable and lead to the development of inhomogeneities. In Section~\ref{sec:inhomogeneous_spheres} we will exploit this feature  to find static inhomogeneous solutions with spherical symmetry in the gauge theory directions, which we will interpret as critical bubbles. Then in Section~\ref{sec:thermodynamics} we explain how to extract their thermodynamic properties.

	\subsection{Homogeneous solutions}
	\label{sec:homogeneous}
	Homogeneous, finite-temperature states in the gauge theory correspond to bulk solutions with a homogeneous horizon. We will construct these solutions in this section. In addition to allowing us to determine the phase diagram of the theory, these solutions will later provide a seed for  the construction of their inhomogeneous counterparts.
	
	A convenient ansatz to find homogeneous black brane solutions is
	\begin{equation}\label{eq:static}
		\dd s^ 2 = \frac{L^ 2}{z^2}\left[ -f(z)\dd t^ 2 + g(z)\parent{\dd\rho^2 + \rho^2 \dd\Omega_{(2)}} + \dd z^ 2 
		\right]\,,
	\end{equation}
	with $f(z), g(z)$ functions of the holographic coordinate to be determined, and 
	\begin{equation}
		\dd\Omega_{(2)} = \dd\theta^2_1 + \sin^2\theta_1 \dd\theta^2_2 
	\end{equation}
	the line element of a two-dimensional sphere. The boundary is located at $z=0$. The scalar $\phi$ is also chosen to depend only on the holographic coordinate, namely: 
	\begin{equation}
		\phi = \homphi(z) \,.    
	\end{equation}
	
	The metric in Eq.~\eqref{eq:static} is of the form
	\begin{equation}\label{eq:def:FG}
		\dd s^2 = \frac{L^2}{\tilde z^2}\parent{\dd \tilde z^2 + g_{\mu\nu}(\tilde z,x)\dd x^\mu\dd x^\nu}
	\end{equation}
	with $\tilde z=z$, which, in particular, has $G_{\mu \tilde z} = 0$, and is thus  written in  Fefferman-Graham (FG) coordinates. The condition for a regular horizon at some position $z=\zH$ is that $f(z)$ has a double zero and that the rest of the  functions are finite:
	\begin{equation}
		\label{eq:IR_expansion}
		f(z) = \fH (z-\zH)^2+ \cdots\,,\qquad
		g(z) = \gH +\cdots\,,\qquad
		\homphi(z) = \phi_\text{\tiny H}+ \cdots\,.
	\end{equation}
	For later purposes, note that $\fH$ has dimensions of mass squared. We can use this series expansion to seed a numerical integrator such as Mathematica \verb|NDSolve| to find the different profiles for the functions all the way up to the boundary at $z=0$, where the metric asymptotes to pure AdS. In this limit,  the functions $f$ and $g$ approach constant values, which can be set to one taking advantage of the three independent rescaling symmetries (see Ref.~\cite{Faedo:2024zib} for details)
	\begin{equation}
		\label{resca}
		g \to \mu_g^2 \,g \,, \qquad f \to \mu_f^2 \,f
		\,, \qquad z \to \mu _z \,z \,.
	\end{equation} 
	With this choice, the expansions near the boundary become
	\begin{eqnarray}
		\label{UVexp_alt}
		\begin{aligned}
			\homphi(z) &= \source z + \fv z^3 + O(z^5)\,,
			\\[2mm]
			f(z) &= 1 - \frac{\source^2}{3} z^2 + f_4 z^4 + O(z^6)\,,\\[2mm]
			g(z) &= 1 - \frac{\source^2}{3} z^2 +\frac{1}{27} \Big[ 2 \left(\source^4-9 \source \fv\right)-9 f_4\Big] z^4 + O(z^6)\,.\\
		\end{aligned}
	\end{eqnarray}
	The coefficient $\source$ in the leading fall-off of the scalar has dimensions of mass. In the dual boundary theory it plays the role of a coupling or source for the scalar operator $\mathcal{O}$ dual to~$\phi$. In general, a solution with a given value of $\phi_s$ corresponds to a theory with a coupling~$\fs$ responsible for the breaking of conformal invariance, and a constant source $J$, in such a way that $\phi_s=\fs+J$. We split $\phi_s$ in this way because we will think of $\fs$ as the scale defining a given theory, and of $J$ as a quantity that we introduce in order to compute the generating functional of correlation functions of $\mathcal{O}$ in that theory. We will set $J=0$ until Section~\ref{sec:effective}, where we will reintroduce $J$ in order to compute the effective action. The expressions for non-zero $J$ can be obtained by replacing $\fs\to \fs+J$ in all the formulas below. We will see below that the coefficient of the subleading term, $\fv$, is related to the expectation value of $\mathcal{O}$. Similarly, the coefficient $f_4$ is related to the energy density and the pressures in the gauge theory. 
	
	The first rescaling in Eq.~\eqref{resca} implies that there is a radially conserved quantity
	\begin{equation}\label{eq:B_radial_quantity}
		Q \equiv \frac{L^3}{z^3}\sqrt{\frac{g(z)}{f(z)}}  \Big[g(z)f'(z) - f(z)g'(z)\Big]\,,
	\end{equation}
	meaning that \(\partial_z Q = 0\).\footnote{In contrast, the conserved quantities under the other two rescalings are identically zero.} 
	This radial conservation law may be used to partially relate UV and IR quantities. More precisely, $Q$ can be expressed in terms of $f_4$, $\fs$ and $\fv$ using the the UV expansions in Eq.~\eqref{UVexp_alt}, while evaluating $Q$ using the IR asymptotics in Eq.~\eqref{eq:IR_expansion} allows us to express it in terms of \(\fH\) and \(\gH\). From this we conclude that
	\begin{equation}
		f_4 = -\frac{3 \fH^{\frac{1}{2}} \gH^{\frac{3}{2}} }{8 \zH^3} + \frac{1}{18} (\fs^4 - 9 \fs \fv)\,.
		\label{eq:f4_equation}
	\end{equation}
	
	\subsection{Inhomogeneous solutions}
	\label{sec:inhomogeneous_spheres}
	Our goal in this section is to construct inhomogeneous solutions corresponding to  critical bubbles in the gauge theory. In the gauge theory, these configurations are spherically symmetric and hence their profile depends only on the radial direction $\rho$. The corresponding solutions in the bulk  will depend on $\rho$ and on the holographic coordinate $z$. 
	
	To construct inhomogeneous solutions, we use DeTurck's trick, see Refs.~\cite{Headrick:2009pv,Wiseman:2011by,Dias:2015nua}. This is a useful procedure to fix the gauge in such a way that Einstein's equations are elliptic. The trick relies on choosing a reference metric $\overline{G}_{MN}$ which behaves near the boundary as the metric we are seeking, ${G}_{MN}$. Then one defines the \textit{DeTurck vector},
	\begin{equation}\label{eq:DeTurck_vector}
		\xi^P = G^{MN}(\Gamma_{MN}^P- \overline{\Gamma}_{MN}^P)\,,
	\end{equation}
	where $\Gamma_{MN}^P$ and $\overline{\Gamma}_{MN}^P$ are the Christoffel symbols of the metrics $G_{MN}$ and $\overline{G}_{MN}$, respectively. We then look  for solutions to 
	\begin{equation}\label{eq:EE_with_DeTurck}
		R_{MN} - \frac{R}{2}G_{MN}-\parent{\nabla_{(M}\xi_{N)} - \frac{1}{2}\nabla_{A}\xi^{A}G_{MN}} =  T_{MN}\,,
	\end{equation}
	where $T_{MN}$ is the stress tensor of the scalar field. When $\xi=0$, Eq.~\eqref{eq:EE_with_DeTurck} reduces to the Einstein equations derived from Eq.~\eqref{eq:actionGH}. Potential solutions with $\xi\neq 0$ are called \textit{DeTurck solitons} and it can be proven that they do not exist under certain conditions, see Refs.~\cite{Figueras:2011va,Figueras:2016nmo}. Here, we monitor $\xi_A\xi^A$ and check that $\xi_A$ vanishes in our solutions. 
	
	Let us now introduce our choice of reference metric $\overline{G}_{MN}$. A usual choice in this context is to use a black brane in pure AdS, which in terms of Eq.~\eqref{eq:static} reads
	\begin{equation}
		f(z) = 1+\frac{z^4}{\zH^4} -\frac{4z^4}{z^4+\zH^4},\qquad  g(z)=1+\frac{z^4}{\zH^4}\,.
		\label{not}
	\end{equation}
	In our case we have made a different choice, because the solutions we seek will asymptote  at large distances (at large values of $\rho$) to one of the homogeneous solutions constructed in Section~\ref{sec:homogeneous}, which differ from Eq.~\eqref{not} because of the non-zero scalar field profile. The physical reason for this is that the bubble configuration is a localized perturbation within a given metastable solution. 
	Consequently, for each choice of temperature, we take the homogeneous metastable solution at that temperature as the reference metric $\overline{G}_{MN}$. One may be concerned that this makes our reference metric already numerical. However, this will not pose a problem because we will be able to determine it with very high precision. Moreover, with this choice of reference metric all the components of the DeTurck vector except $\xi^z$ and $\xi^\rho$ vanish identically in Eq. \eqref{eq:DeTurck_vector}. Then $\xi^A$ is spacelike and $\xi^A\xi_A= 0$ if and only if $\xi^A=0$.
	
	To find inhomogeneous, spherically symmetric solutions we start with the following ansatz for the metric:
	\begin{equation}
		\label{eq:static_PDE}
		\begin{aligned}
			\dd s^ 2 = \frac{L^ 2}{z^2}\Big[-f(z)\Qtt(\rho,z)\dd t^ 2 + g(z)\parent{\Qxx(\rho,z)\dd\rho^2 + \rho^2 \Qyy(\rho,z) \dd\Omega_{(2)}}& \\\qquad\qquad+ \Qzz(\rho,z)\dd z^ 2 + (\zH - z) \Qzx(\rho,z)\dd\rho \,\dd z & \ \Big]\,,
		\end{aligned}
	\end{equation}
	together with
	\begin{equation}
		\phi = \homphi(z) \Qphi(\rho,z)\,.
	\end{equation}
	Note that there is no available gauge freedom  to eliminate the off-diagonal, $\dd\rho\, \dd z$-term in the metric \eqn{eq:static_PDE}, because the gauge is fixed by the choice of reference metric, which assigns a physical meaning to the $\rho$ and $z$ coordinates. In other words, in the context of the De Turck trick the condition $\xi^P = 0$ should be seen as fixing a coordinate gauge, and in this gauge the metric takes the form \eqref{eq:static_PDE}. Because of this, the metric \eqref{eq:static_PDE} is not in FG form; we will return to this point below. 
	
	For each temperature, the functions $f(z)$, $g(z)$ and $\homphi(z)$ in Eq.~\eqref{eq:static_PDE} are taken to be those of the homogeneous metastable solution approached as $\rho\to\infty$, which we chose as the reference metric. Hence, $Q_i(\rho,z)$ parametrize the deviation from this solution. Since this deviation must vanish far away from the bubble location, we impose the boundary conditions
	\begin{equation}
		\lim_{\rho\to\infty} Q_i(\rho,z) = 1 \quad \text{for }i = 0,\cdots, 4\,,\qquad \lim_{\rho\to\infty} Q_5(\rho,z) = 0\,.
	\end{equation}
	In addition, we require regularity at $\rho = 0$.
	
	Regarding the boundary conditions in the holographic direction, a near-boundary analysis shows that the different functions must approach the boundary as 
	\begin{equation}
		\label{eq:UV_expansion_inhomogeneous}
		\begin{aligned}
			\Qtt(\rho,z)&=  1+\qtt(\rho) z^4 + O(z^6\log z)\,, \\[\spacelist]
			\Qxx(\rho,z)&=  1 + \qxx(\rho) z^4 + O(z^6\log z)\,,    \\[\spacelist]
			\Qyy(\rho,z)&=  1 - \frac{1}{2} \left( 4 \fs^2 \, \ftwo(\rho) +\qtt(\rho) + \qxx(\rho) \right) z^4 + \mathcal{O}(z^6\log z)\,,
			\\[\spacelist]
			\Qzz(\rho,z)&= 1 + 2 \fs^2 \ftwo(\rho) z^4 + O(z^6\log z)\,,  \\[\spacelist]
			\Qzx(\rho,z)&=  \qfive(\rho) z^5 + \frac{\fs^2}{2\zH} \ftwo'(\rho) z^5\log z +  O(z^6\log z)\,, \\[\spacelist]
			\Qphi(\rho,z)&= 1 + \ftwo(\rho) z^2 +  O(z^4)\,.
		\end{aligned}
	\end{equation}
	The appearance of logarithms in these formulas may be surprising given that there are no conformal anomalies in the boundary theory under consideration. The reason is that the metric \eqref{eq:static_PDE} is not written in FG form, since $G_{\rho z}\neq 0$. This point will become clear below. We see that, in addition to the parameters of the chosen background solution, new undetermined functions of the radial coordinate appear in the expansion: $\qtt(\rho)$, $\qxx(\rho)$, $\ftwo(\rho)$ and $\qfive(\rho)$. These will account for the inhomogeneities of the different thermodynamic quantities, as we explain in the next Section. 
	
	Regularity at the horizon imposes the following boundary conditions:
	\begin{equation}
		\Qtt(\rho,\zH) = \Qzz(\rho,\zH)\,,\quad\partial_z \Qzx(\rho,\zH)-\frac{\Qzx(\rho,\zH)}{2 \zH} =0\,,\quad \partial_z Q_i(\rho,\zH) = 0\ \mbox{\, for  \,} i= 1,2,3\,.
	\end{equation}
	The first condition ensures that the temperature is constant along the horizon despite its inhomogeneous nature. 
	
	If we evaluate the norm of $\xi$ on the asymptotic expansion shown in Eq.~\eqref{eq:UV_expansion_inhomogeneous}, we obtain that $\xi_A\xi^A=0$ implies the following relation between the coefficients:
	\begin{equation}\label{eq:conservation_from_xi}
		\qxx'(\rho) = -\frac{1}{\rho}
		\parent{
			4 \fs^2 \, \ftwo(\rho) 
			+  \qtt(\rho) 
			+ 3 \qxx(\rho)} - \frac{3}{2}\fs^2  \ftwo'(\rho)
		\,.
	\end{equation}
	In the next Section we will see that this equation is in fact equivalent to the conservation of the stress tensor in the boundary QFT.
	
	Below we will describe in detail the solutions to the Einstein-dilaton equations with these boundary conditions. Each solution can be constructed from the previous one by slightly changing the temperature and letting the profiles of the different functions relax to a new solution, via a Newton-Raphson procedure. One of the most challenging steps in these types of problems is to get the first non-trivial solution. To construct it, we started with a homogeneous solution close to the turning point, namely to the lowest-temperature state on the metastable branch (see Fig.~\ref{fig:phase_diagram}). At this point  we expect the bubble amplitude to be small, i.e.~we expect the bubble solution to be everywhere close to the homogeneous state. Then we seeded the Newton--Raphson algorithm with the homogeneous metastable solution plus a small convenient perturbation. After a few trials and errors the code converged to a new, inhomogeneous solution. Details of the computations can be found in Appendix~\ref{app:implementation}.

	\subsection{Thermodynamics}
	\label{sec:thermodynamics}
	Once the numerical solutions for the critical bubbles are constructed, we can compute their thermodynamic properties. As usual, the entropy of each solution can be computed from the area of the horizon:
	\begin{equation}
		\label{eq:entropy_bubble}
		\begin{aligned}
			S  &= \frac{2\pi}{\kappa_5^2}\text{Area}(z=\zH) 
			\\[2mm]
			&= \frac{2 L^3}{\kappa_5^2} \frac{ \pi\gH^{3/2}}{\zH^3} \times 4\pi \int_0^\infty\dd\rho \, \rho^2 \Qxx(\rho, \zH)^{1/2} \Qyy(\rho, \zH)
			\\[2mm]
			&\equiv 4\pi\int_0^\infty\dd\rho\, \rho^2 s(\rho)\,,
		\end{aligned}
	\end{equation}
	where the last identity defines what we mean by the entropy density per unit physical volume in the boundary theory---we will come back to this point at the end of this section.  In Eq.~\eqref{eq:entropy_bubble}, the 
	term on the left of the ``$\times$'' is the entropy density of the metastable, homogeneous solution to which the bubble solution asymptotes as $\rho \to \infty$. In the homogeneous case $\Qxx=\Qyy=1$ and the term on the right of the ``$\times$'' gives simply the volume of flat space, which is of course infinite. However, the difference between the entropy of the bubble solution, \eqref{eq:entropy_bubble}, and that of the corresponding homogeneous (metastable) one, $S_+$, is given by 
	\begin{equation}
		\Delta S = S-S_+  = \frac{2 L^3}{\kappa_5^2} \frac{ \pi \gH^{3/2}}{\zH^3} \times 4\pi \int_0^\infty\dd\rho \, \rho^2 \parent{\Qxx(\rho, \zH)^{1/2} \Qyy(\rho, \zH)-1}\,,
	\end{equation}
	and is finite. Following Ref.~\cite{Laine:2016hma}, we will use an analogous notation, $\Delta X = X - X_{+}$, for any quantity $X$. 
	
	To compute the temperature, we note that the horizon is generated by the Killing vector field 
	\begin{equation}
		\label{eq:Killing_on_Horizon}
		\zeta = \partial_t\,,
	\end{equation}
	which is timelike for $z<\zH$ and null at $z=\zH$, i.e.~$\zeta_M\zeta^M|_{z=\zH} = 0$. The surface gravity and the temperature are thus given by 
	\begin{equation}
		\kH^ 2 = -\frac{1}{2}(\nabla^M\zeta^N)(\nabla_M\zeta_N)\Big|_{z=\zH} = \fH
	\end{equation}
	and  
	\begin{equation}
		T = \frac{\kH}{2\pi} = \frac{\fH^{1/2}}{2\pi}\,,
	\end{equation}
	respectively. Recall that $\fH$ has units of energy squared---see Eq.~\eqref{eq:IR_expansion}. Note that, as expected, the temperature is constant all along the horizon despite the inhomogeneous nature of the latter, and it coincides with the temperature of the metastable, homogeneous state to which the bubble solution asymptotes. Physically, this property is  a consequence of the 0th law of black hole mechanics.
	
	To obtain the remaining  properties of the critical bubbles we use the well-known holographic renormalization procedure \cite{Henningson:1998gx,Balasubramanian:1999re,deHaro:2000vlm,Skenderis:2002wp}, whose details we discuss in Appendix~\ref{app:holoren}. 
	To implement this procedure,  it is convenient to rewrite the metric in FG coordinates. In other words, we perform a change of variables $\zt = \zt(\rho,z)$, $\rhot = \rhot(\rho, z)$ such that the metric \eqref{eq:static_PDE} in the tilde coordinates takes the form \eqref{eq:def:FG}. Only the coordinate transformation near the boundary is needed, which takes the form
	\begin{equation}
		\label{eq:change_of_variables}
		\begin{aligned}
			\zt &= z + \frac{1}{4}\fs \ftwo(\rho)z^5 +O(z^7\log z)\,,\\[2mm]
			\rhot &= \rho + \frac{1}{36} \left(6 \zH\, \qfive(\rho) 
			- 2 \fs^2\, \ftwo'(\rho) 
			+ 3 \fs^2  \ftwo'(\rho) \log z\right) z^6 + O(z^ 8\log(z))\,.
		\end{aligned}
	\end{equation}
	Note that $\rho$ and $\tilde\rho$ coincide at the boundary, i.e.~at $z=0$. For this reason we can use them interchangeably  in the UV quantities below that do not depend on $\zt$.\footnote{This would not apply to functions evaluated at the horizon.} The near boundary expansion in terms of the FG coordinates becomes
	\begin{eqnarray}\label{eq:UV_general_inhomogeneousFG}
		{g}_{\mu\nu}(\zt,x)&=&\frac{L^2}{\zt^2} \parent{\eta_{\mn} \,+\,  \frac{\fs^2}{3} \eta_{\mn}\zt^2\,+\, \gamma_{(4)\mn}(\rho)  \zt^4  \,+\, O(\zt^6) }\,,\nonumber\\[\spacelist]
		\phi(\zt,x)&=& \zt\parent{\fs+ \phi_{(2)}(\rho)\zt^2 +O(\zt^4)}\,,
	\end{eqnarray}
	where now $\gamma_{(4)\mn}(\rho)$ and $ \phi_{(2)}(\rho)$ are combinations of the undetermined UV parameters appearing in Eqs.~\eqref{UVexp_alt} and \eqref{eq:UV_expansion_inhomogeneous}. In contrast to Eq.~\eqref{eq:UV_expansion_inhomogeneous} (see also Appendix \ref{app:holoren}) there are no logarithms in the FG form of the metric or the scalar field, consistently with the fact that there are no conformal anomalies in the boundary theory. The logarithms appear in the change of variables \eqref{eq:change_of_variables} from the FG coordinates $\rhot, \zt$ to those in Eq.~\eqref{eq:static_PDE}.
	
	Once the metric is expressed in FG coordinates, it is  traightforward to obtain an expression for the  expectation value of the energy-momentum tensor, which in our (spherical) coordinates reads
	\begin{equation}\label{eq:EM.Spherical}
		\left\langle {(T^{\text{\tiny{QFT}}}) }^{\mu}_{\ \nu}\right\rangle = \text{diag}\Big( - e(\rho),\,  \PL(\rho),\PT(\rho),\,   \PT(\rho) \, \Big)\,.
	\end{equation} 
	Here $e(\rho)$ is the energy density, and $\PL$ and $\PT$ are the longitudinal and transverse pressures with respect to the $\partial_\rho$ direction, respectively. The general formula is given in Eq.~\eqref{eq:EMT_app}, but there is an enormous simplification in our case because the boundary metric is flat and the source of the scalar operator is constant. Under these circumstances
	\begin{equation}
		\label{eq:EMtensor}
		\begin{aligned}
			\left\langle T^{\text{\tiny{QFT}}}_{\mu \nu}\right\rangle &= \frac{2L^3}{\kappa_5^2}
			\parent{\gamma_{(4)\mu\nu}(\rho) + \eta_{\mu\nu}\parent{\fs \phi_{(2)}(\rho) + \parent{{\be} - \frac{1}{18}}\fs^4}}\,,
		\end{aligned}
	\end{equation}
	where $\be$ is a finite counterterm encoding the renormalization-scheme ambiguity of the model, see Eq.~\eqref{eq:ct_app}. Unless otherwise stated, we set $\be=\lambda_4$, so that the zero-temperature limiting solution has vanishing energy. From this we conclude that the total energy is
	\begin{equation}
		\label{eq:expression_energy}
		\begin{aligned}
			E &=  4\pi \int_0^\infty \dd\rho \, \rho^2\, e(\rho)\\[2mm]
			& = \frac{2 L^3}{\kappa_5^2}\times 4\pi \int_0^\infty \dd\rho \, \rho^2\left(-f_4 + \frac{\fs^4}{18} - \be  \fs^4 - \fs \fv - \frac{3}{2} \fs^2 \ftwo(\rho) - \qtt(\rho)\right) \,,
		\end{aligned}\,
	\end{equation}
	while the pressures read
	\begin{equation}\label{eq:expression_pressures}
		\begin{aligned}
			\PL &= \frac{2 L^3}{\kappa_5^2}\left(-\frac{f_4}{3} + \frac{\fs^4}{54} + \be  \fs^4 + \frac{\fs \fv}{3} + \frac{3}{2} \fs^2 \ftwo(\rho) + \qxx(\rho)\right)\,,\\[\spacelist]
			\PT &= \frac{2 L^3}{\kappa_5^2} \left(- \frac{f_4}{3} + \frac{\fs^4}{54} + \be \fs^4 + \frac{\fs  \fv }{3} + \frac{1}{2} \parent{-\fs^2 \ftwo(\rho) -\qtt(\rho) - \qxx(\rho)}\right) \,.\\[\spacelist]
		\end{aligned}
	\end{equation}
	
	Notice that, as for the entropy density, the corresponding homogeneous expressions are obtained by setting the inhomogeneity to zero, $\qtt(\rho)=\qxx(\rho)=\ftwo(\rho) = 0$, in which  case $\PL = \PT\equiv p$. The energy density and the pressure of the homogeneous solutions read
	\begin{equation}
		\label{eq:homogeneous_E_P}
		e= \frac{2 L^3}{\kappa_5^2} \left( -f_4 + \frac{\fs^4}{18} - \be  \fs^4 - \fs \fv \right),\qquad
		p= \frac{2 L^3}{\kappa_5^2}\left( -\frac{f_4}{3} + \frac{\fs^4}{54} + \be \fs^4 + \frac{\fs \fv}{3}\right),
	\end{equation}
	respectively.
	
	In analogy with the entropy, we consider the energy difference between the inhomogeneous  and the homogeneous solutions. This takes the form
	\begin{equation}
		\Delta E \equiv  E - E_{+}= \frac{2\pi L^3}{\kappa_5^2}\times 4\pi \int_0^\infty \dd\rho \, \rho^2\left( \frac{5}{2} \fs^2 \ftwo(\rho) + 3 \qxx(\rho) + \rho \, \qxx'(\rho)\right) <0\,,
	\end{equation}
	and  is finite. The negative sign of $\Delta E$ is due to the fact that the bubble replaces a volume of  high-energy phase by one of low-energy phase.  With all this information we can also find the free energy as
	\begin{equation}\label{eq:free_energy1}
		F = E - TS \,.
	\end{equation}
	The free energy difference between the critical bubble and the homogeneous metastable state is therefore  
	\begin{equation}\label{eq:free_energy2}
		\Delta F \equiv F - F_+ = \Delta E - T\Delta S\,.
	\end{equation}
	This  will play a crucial role below, since it determines the nucleation rate of the critical bubbles. In Appendix \ref{app:free energy} we show that $F/T$ coincides with the on-shell Euclidean action, as expected. 
	
	Holographic renormalization also provides an expression for the expectation value of the scalar operator in terms of the asymptotic data, Eq.~\eqref{eq:VEV_app}, which in our case becomes
	\begin{equation}
		\label{eq:formula_VEV}
		\VEV = 
		\frac{2L^3}{\kappa_5^2}\parent{
			{-2\phi_{(2)}(\rho)-4\be \fs^ 3}
		}=
		\frac{2L^3}{\kappa_5^2}\parent{
			{- 2\ftwo(\rho) -2\fv -4\be \fs^ 3}
		}\,.
	\end{equation}
	
	Note the appearance of Newton's constant in front of all the quantities above. To translate it to a gauge theory quantity, we will use the coefficient of the Euler density appearing in the Weyl anomaly of the dual field
	theory (see Ref.~\cite{Henningson:1998gx}),
	\begin{equation}\label{eq:Nsquared}
		a = \frac{4L^ 3}{\kappa_5^ 2}\propto N^2\,.
	\end{equation}
	Here we have indicated explicitly the fact that, in top-down scenarios, this quantity is proportional to the number of degrees of freedom in the dual QFT. 
	
	We close this section with an observation regarding the densities of thermodynamic quantities for inhomogeneous solutions. The energy density and the pressure are always unambiguously defined at any spacetime point because they can be extracted from the stress tensor. The temperature is also well defined because the 0th law of black hole mechanics implies that it is constant along the horizon. In contrast, defining the entropy density requires the choice of a map between points at the boundary and points at the horizon, and this is not unique. 
	In Eq.~\eqref{eq:entropy_bubble} we chose the map suggested by the gauge in which the metric takes the form  \eqn{eq:static_PDE}. 
	From the entropy and the energy densities we then define the free energy density through the usual relation
	\begin{equation}
		f=e-Ts \,.
		\label{freedensity}
	\end{equation}
	Needless to say, the free energy density suffers from the same ambiguities as the entropy density. To illustrate this, we note that we would arrive at a different definition if we used the tilde coordinates in Eq.~\eqn{eq:change_of_variables} in which the metric takes the FG form. A third natural choice would be to map the points at the boundary to points at the horizon by shooting in-going null geodesics from the boundary, as suggested by the  fluid/gravity correspondence~\cite{Bhattacharyya:2007vjd}. Crucially, the physical results are independent of these choices, since they depend only on integrated quantities (such as the nucleation rate) or can be expressed in terms of well-defined observables, such as the energy density (for instance, the bubble wall profile).

	\section{Results}\label{sec:results}
	
	\subsection{Phase diagram}
\begin{figure}[t]
		\centering
		\includegraphics[width=0.80\linewidth]{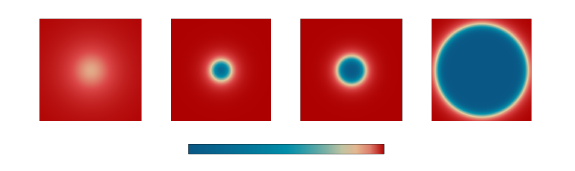}
		\put(-320,95){\footnotesize \color{color1} $\displaystyle \ratioT = 2.2\times10^{-3}$}
		\put(-230,95){\footnotesize \color{color2}
			$\displaystyle \ratioT = 0.20$}
		\put(-152,95){\footnotesize \color{color3}
			$\displaystyle \ratioT = 0.44$}
		\put(-75,95){\footnotesize \color{color4}
			$\displaystyle \ratioT = 0.84$}
		\put(-190,-8){\footnotesize 
			$\displaystyle \frac{e(\rho) - e_-}{e_+ - e_-}$}
		\put(-115,0){\footnotesize $1$}
		\put(-235,0){\footnotesize $0$}
		\vspace{7mm}
		\includegraphics[width=0.80\linewidth]{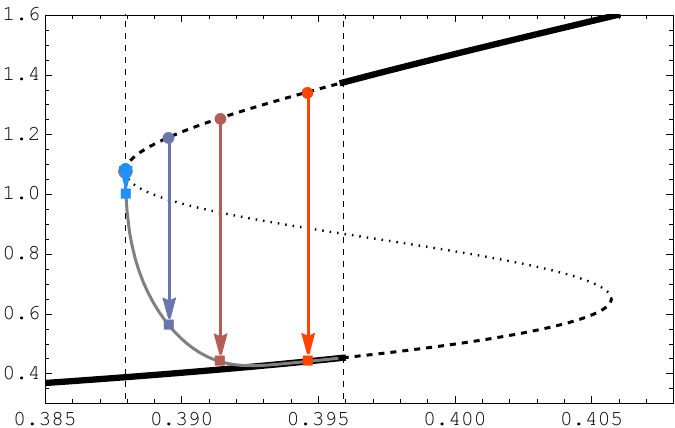}
		\put(-370,120){\small $\displaystyle \frac{2e}{a\fs^4}$}
		\put(-180,-15){\small $T/\fs$}
		\put(-182,215){\footnotesize $T=T_c$}
		\put(-295,215){\footnotesize $T=T_0$}
		\caption{\small 
			Energy density as a function of temperature  for the theory with $\lambda_4 = -1/4$, $\lambda_6 = 1/10$. The thick solid curves correspond to the two branches of stable configurations. The two vertical, thin, dashed black lines indicate the location of the turning point at $T = T_0$, where the spinodal branch begins, and the first-order phase transition at $T = \Tc$. The two dashed curves that extend between the critical temperature and the corresponding turning point are metastable branches. They are connected by the spinodal branch (intermediate dotted curve), where the system is both thermodynamically and dynamically unstable. The gray solid curve that interpolates between $T_0$ and $\Tc$ indicates the energy density at the center of the critical bubbles. At the top, we present the density plot for the relative energy density of four representative bubble solutions, indicating the value of their relative temperature difference $\ratioT$, defined in Eq.~\eqref{eq:rel_temperature_difference}. Their energy profiles expand from the metastable branch (dots) to the corresponding value at the center of the bubble (squares), as indicated by the arrows. In the legend, $e_+$ and $e_-$ refer to the energy densities in the metastable, high-energy phase and in the stable, low-energy phase, respectively.}
		\label{fig:phase_diagram}
	\end{figure}

\begin{figure}[t]
		\centering
		\includegraphics[width=0.8\linewidth]{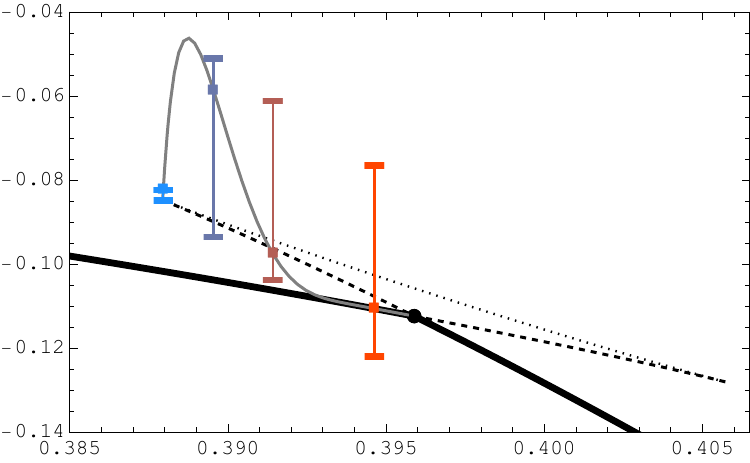}
		\put(-370,100){$\displaystyle \frac{2f}{a\fs^4}$}
		\put(-170,-15){$T/\fs$}
		\\\vspace{6mm}
		\includegraphics[width=0.8\linewidth]{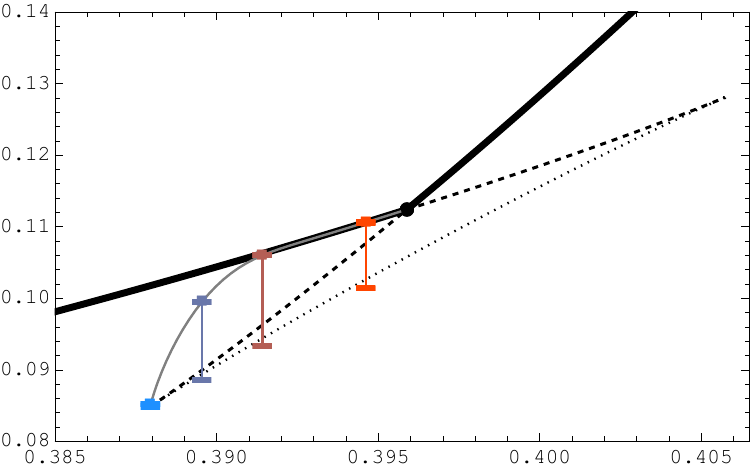} 
		\put(-370,100){$\displaystyle \frac{2p}{a\fs^4}$}
		\put(-170,-15){$T/\fs$}
		\caption{\small Free energy density defined through \eqn{freedensity} (top) and pressure (bottom) of the homogeneous solutions as  functions of the temperature. The conventions are the same as in Fig.~\ref{fig:phase_diagram}, with the thickest curves representing homogeneous, stable states. The critical temperature, marked by a thick black dot, is  the point where these curves cross each other. For the same four bubbles shown in Fig.~\ref{fig:phase_diagram}, we have drawn bars indicating the range probed by each bubble (see also Fig.~\ref{fig:results_bubbles}), extending from the origin (marked by a square) to infinity (on the metastable dashed branch). Close to $T_c$, the state at the bubble center lies near the stable homogeneous state, although the bubble wall itself remains non-trivial. Note that $p=-f$ for homogeneous states but not for inhomogeneous ones, as expected on general grounds. }
		\label{fig:free_energy}
	\end{figure}

	Let us now show our results. For concreteness, we will restrict our analysis to the case $\lambda_4=-1/4$ and \mbox{$\lambda_6=1/10$}. We expect qualitatively  similar  results for other values for which the theory also exhibits a FOPT.  Once these parameters are specified, we can obtain a family of homogeneous black brane solutions parametrized by the value of the scalar $\phi$ at the horizon, $\phi_{\text{\tiny H}}$. Some homogeneous solutions, with different values of $\phi_{\text{\tiny H}}$, share the same value of the temperature in units of the source, $T/\fs$. This leads to the mutivaluedness visible in  Fig.~\ref{fig:phase_diagram}, which is characteristic of a FOPT. In the mutivalued region, the thermodynamically preferred phase (in the canonical ensemble) is that with the lowest free energy. In Fig.~\ref{fig:free_energy} (top), where we show the free energy density as a function of the temperature, we see that the system undergoes a FOPT at the critical temperature $T=\Tc\simeq 0.3959\fs$, where the curve  crosses itself. Since for homogeneous solutions the free energy equals minus the pressure, it is clear that the latter is continuous across the FOPT, while the energy density jumps by almost a factor of 3. This also implies that below $T = T_c$ the stable phase has a higher pressure than the metastable phase, even though it has a lower energy density---see Fig.~\ref{fig:free_energy} (bottom).
	
	Below $\Tc$ there is a range of temperatures $T\in(T_0,\Tc)$, with $T_0\simeq 0.3879 \Lambda$, where a \textit{supercooled} metastable branch of states exists. 
	We will find it useful to define the \textit{relative temperature difference} as
	\begin{equation}\label{eq:rel_temperature_difference}
		\ratioT(T) = \frac{T-T_0}{T_c-T_0}\,.
	\end{equation}
	This quantity takes values between zero (at $T_0$) and one (at $T_c$), and hence it is helpful to indicate the position of each solution in the metastable branch. Note that $T_0$ is very close to $T_c$, since 
	\begin{equation}
		\frac{T_c-T_0}{T_c}\simeq 2.01 \times 10^{-2}\,.
		\label{Tclose}
	\end{equation}
	Similarly, a \textit{superheated} metastable branch exists above $\Tc$. The homogeneous states on these branches have higher free energy than those in the preferred phases but they are dynamically stable against small perturbations. Whether a perturbation triggers the phase transition depends on its size, as will be clear from our analysis of critical bubbles.

	\subsection{Critical bubbles}\label{sec:Critical_Bubbles}
	Let us discuss the solutions and properties of the critical bubbles, focusing on the metastable supercooled branch. As we anticipated, each of the critical bubbles we construct asymptotes to one of the states on this metastable branch. To construct the first one, we chose a homogeneous state close to the turning point at $T=T_0$ and seeded a Newton-Raphson algorithm with this homogeneous metastable solution plus  a small perturbation. After few trials and errors the code converged to a new, inhomogeneous solution. Once the first bubble is constructed, we vary slightly the value of $T/\fs$ and use the first solution to construct the second one. Then we iterate this process until we cover the full range of temperatures $T\in(T_0, \Tc)$. Details of the computations can be found in Appendix~\ref{app:implementation}.
	
	\begin{figure}[t]
		\hspace{.8cm}
		\includegraphics[width=1\textwidth]{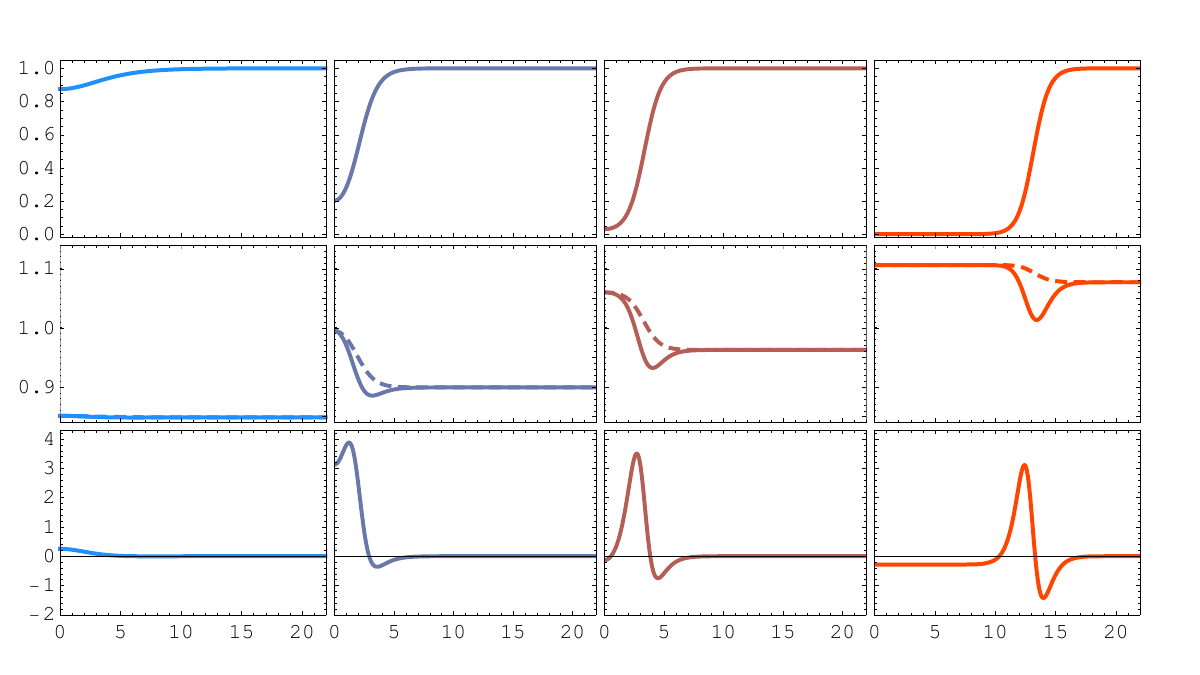}
		\put(-385,225){\footnotesize $\displaystyle \ratioT = 2.2\times 10^{-3}$}
		\put(-280,225){\footnotesize $\displaystyle \ratioT = 0.20$}
		\put(-188,225){\footnotesize $\displaystyle \ratioT = 0.44$}
		\put(-92,225){\footnotesize $\displaystyle \ratioT = 0.84$}
		\put(-467,190){\footnotesize $\displaystyle \frac{e(\rho) - e_-}{e_+ - e_-}$}
		\put(-465,125){\footnotesize $\displaystyle \frac{2p}{a\fs^4}\cdot 10$}
		\put(-460,50){\footnotesize  $\displaystyle \frac{2\Delta f}{a\fs^4}\cdot 10^2$}
		\put(-360,5){\footnotesize $\rho\fs$}
		\put(-265,5){\footnotesize $\rho\fs$}
		\put(-170,5){\footnotesize $\rho\fs$}
		\put(-75,5){\footnotesize $\rho\fs$}
		\caption{\small  
			(Top) Energy density of a bubble as a function of the radial direction, normalized to the difference between the metastable ($+$) and stable ($-$) branches, for the four different choices of $\ratioT$ indicated on each column, where $\ratioT$ is defined in Eq.~\eqref{eq:rel_temperature_difference}. Near $T_0$ (left, $\ratioT = 0$) the profile of the energy density is Gaussian and its characteristic size grows as $T_0$ is approached. On the other side, near $T_c$ (right, $\ratioT = 1$), the profile becomes closer and closer to a hyperbolic tangent. The profiles of the entropy density and the expectation value of $\OO$ exhibit qualitatively similar behavior. (Middle) Transverse (solid curves) and longitudinal (dashed curves) pressures as a function of the radial direction for the same values of the temperature. These coincide at the origin, where isotropicity is restored, and at infinity, where the asymptotic, metastable, homogeneous state is recovered. (Bottom) Free energy density difference, $\Delta f=f(\rho)-f_+$, as a function of the radial direction for the same values of the temperature. These are the profiles that we integrate to obtain the bubble action.
		}
		\label{fig:results_bubbles}
	\end{figure}
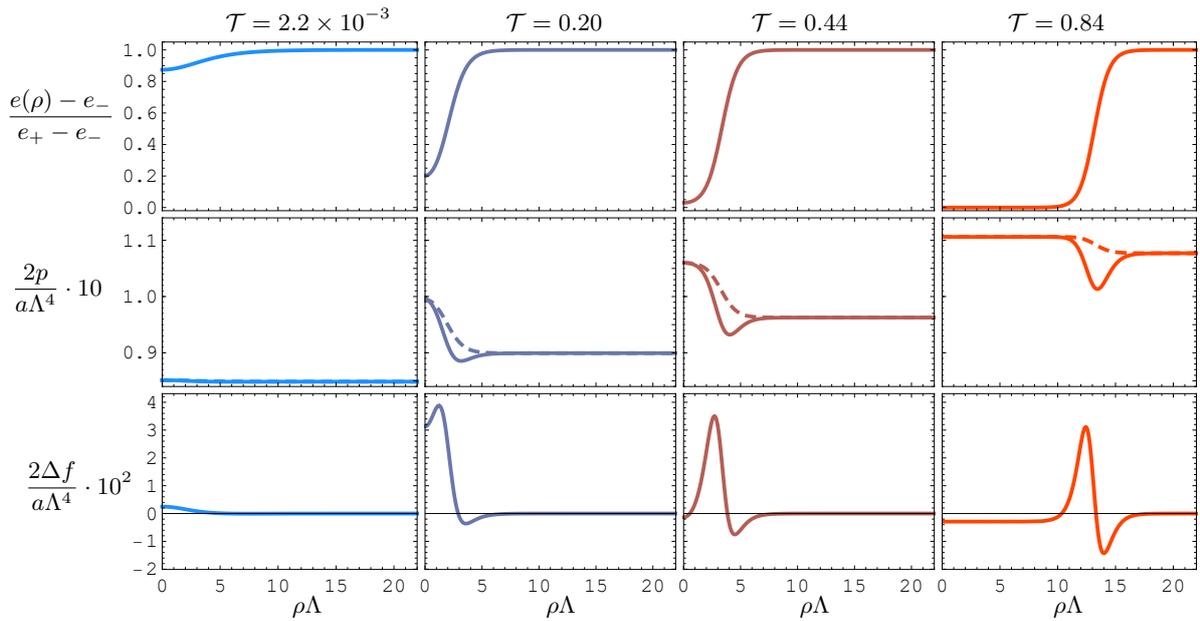
	
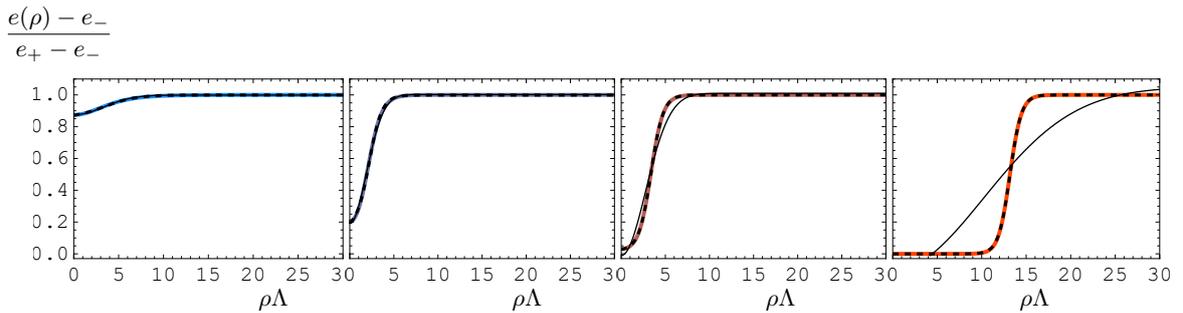
\begin{figure}[!!t]
		\includegraphics[width=1.05\textwidth]{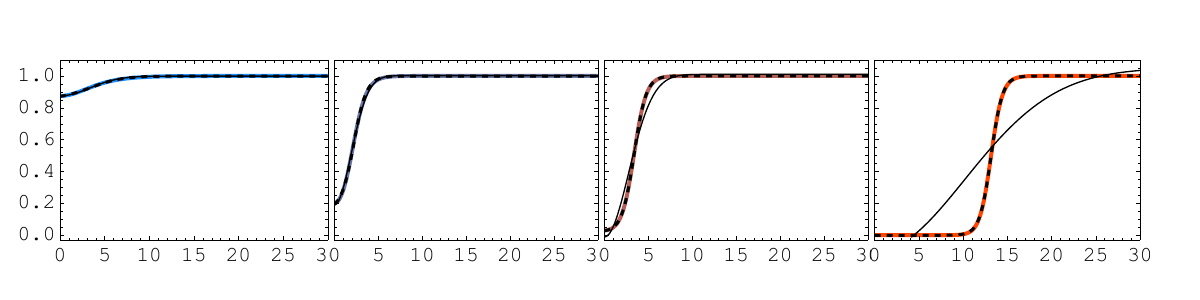}
		\put(-455,105){\footnotesize $\displaystyle \frac{e(\rho) - e_-}{e_+ - e_-}$}
		\put(-360,5){\footnotesize $\rho\fs$}
		\put(-265,5){\footnotesize $\rho\fs$}
		\put(-170,5){\footnotesize $\rho\fs$}
		\put(-75,5){\footnotesize $\rho\fs$}
		\caption{\small Best Gaussian (thin, solid) and hyperbolic tangent (dashed, thick) fits on top of the energy density profile from Fig.~\ref{fig:results_bubbles}~(top). The hyperbolic tangent appears to approximate the profiles well in all the range of temperatures, but it fails to give the correct Newman boundary condition at $\rho = 0$, specially close to $T_0$. The Gaussian profile Eq.~\eqref{eq:Gaussian_profile}, in contrast, has a minimum at the origin, but fails to reproduce the profile away from $T_0$. }
		\label{fig:fits}
	\end{figure}

\begin{figure}[t]
		\centering
		\includegraphics[height=.21\textheight]{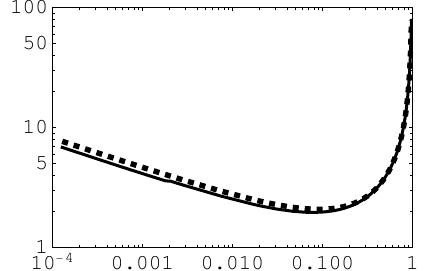}
		\put(-200,145){\small $\radius\fs$}
		\put(-110,-20){\small $\displaystyle \frac{T-T_0}{T_c-T_0}$}
		\hfill
		\includegraphics[height=.21\textheight]{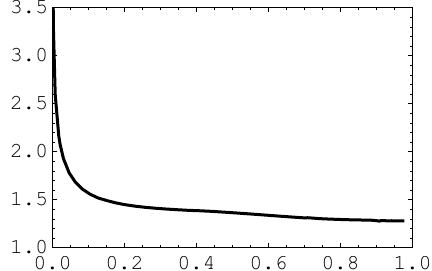}
		\put(-200,145){\small $\lw\fs$}
		\put(-110,-20){\small $\displaystyle \frac{T-T_0}{T_c-T_0}$}
		\caption{\small (Left) Radius of the bubbles from Eq.~\eqref{eq:radius_width1} (solid) and Eq.~\eqref{eq:radius_2} (dashed) as a function of the temperature. (Right)  Width of the bubble wall from Eq.~\eqref{eq:radius_width1} as a function of the temperature.}
		\label{fig:radius_and_wall}
	\end{figure}

	As we vary the temperature and find different solutions, we uncover two qualitatively distinct regimes depending on whether or not $T$ is close to $T_c$, as shown in Fig.~\ref{fig:results_bubbles}. If $T$ is close to  $\Tc$, then the bubbles consist of three parts: the interior of the bubble, where the system is approximately in the low-energy, stable phase; the exterior, which lies  in the high-energy, metastable phase; and a wall that separates these two regions. In particular, the profile of several quantities such as the energy density or the expectation value of the scalar operator become well approximated by a hyperbolic tangent of the form
	\begin{equation}
		\label{eq:radius_width1}
		e (\rho) \simeq  e_0 +  e_m \tanh\parent{\frac{\rho-\radius}{\lw}}\,,
	\end{equation}
	as illustrated in Fig.~\ref{fig:fits}. The bubbles in this regime have a characteristic radius $\radius$ and thickness $\lw$. In order to understand how these quantities depend on the temperature, we fit the profile of the bubbles  to Eq.~\eqref{eq:radius_width1}.  An alternative definition of the size of the bubble can be the value of $\rho =\radius$ where the profile for the energy density acquires its  mean value between the energy density at infinity and at the center of the bubble, i.e.,~by the equation
	\begin{equation}\label{eq:radius_2}
		e\parent{\rho = \radius} = \frac{1}{2}\parent{e(0)+e(\infty) }= \frac{1}{2} \parent{e(0)+e_+}\,. 
	\end{equation}
	We show these two alternative definitions of $\radius$ in  Fig.~\ref{fig:radius_and_wall}, where we see that they essentially coincide. 
	
	As the temperature decreases, the bubble shrinks, and its interior eventually no longer reaches the low-energy (stable) phase.
	This behavior is illustrated in Fig.~\ref{fig:phase_diagram}, where the energy density at the center of the bubble is shown as a gray curve overlaid on the phase diagram. At some point, the profile of the bubble ceases to be well approximated by two homogeneous phases separated by a wall.
	Indeed, close to $T_0$ we reach a second, qualitatively different regime where the profile of different quantities becomes Gaussian, for instance
	\begin{equation}\label{eq:Gaussian_profile}
		e(\rho)\simeq e_0 + e_p \exp\parent{-
			\frac{\rho^2}{2\widthGaussian^2}}\,.
	\end{equation}
	Here we see that as the temperature decreases towards $T_0$ the bubbles grow again and become flatter in the sense that the difference between the energy density at the origin and infinity decreases.
	
	It is worth discussing the pressure profiles for these solutions. As we mentioned, despite the fact that  the solutions are static and obey $\nabla_{\mu}T^{\mu\nu} = 0$, the pressures need not be the same in all directions, since the solutions are anisotropic. In fact, the conservation of the stress tensor reduces to only one non-trivial equation: 
	\begin{equation}\label{eq:conservation}
		0 = \,\, \nabla_\mu T^{\mu \rho}\,\, = \,\, p_{||}'(\rho) +\frac{2}{\rho}\parent{p_{||}(\rho)-p_\perp(\rho)}\,.
	\end{equation}
	As usual, from the bulk perspective this equation is one of the constraints in the Einstein equations evaluated near the boundary. The difference between the longitudinal and transverse pressures is illustrated in Fig.~\ref{fig:results_bubbles}. In particular, we see that  the longitudinal pressure is monotonic, whereas the transverse one is not. Nevertheless, they coincide both at the origin and at infinity, where rotational symmetry requires isotropy.
	
	Interestingly, if we substitute the expression for the pressures \eqref{eq:expression_pressures}  into Eq.~\eqref{eq:conservation}, we find precisely Eq.~\eqref{eq:conservation_from_xi}, which is satisfied on-shell (when $\xi^A = 0$). A way to phrase this result is that DeTurck solitons in this system (i.e. solutions with $\xi^A \neq 0$), if they existed,  would fail to satisfy energy-momentum conservation in the boundary theory. 
	In addition, Eq.~\eqn{eq:conservation} provides a check of our numerics: we can calculate $\PT$ and $\PL$ from the numerical solutions and verify that they indeed satisfy Eq.~\eqref{eq:conservation}.
	
	\subsection{Nucleation rate}
\begin{figure}[t]
		\centering
		\includegraphics[width=0.70\linewidth]{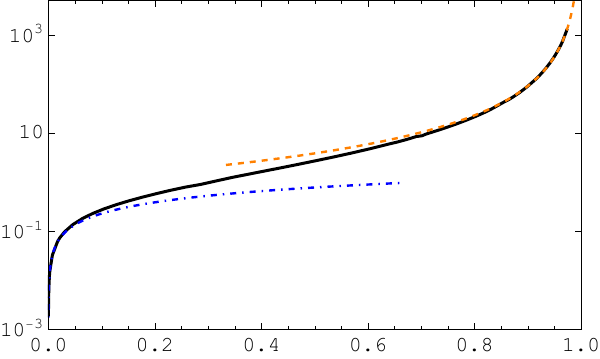}
		\put(-320,100){\small $\displaystyle \frac{2 \Delta F}{ a T}$}
		\put(-160,-15){\small $\displaystyle \frac{T-T_0}{T_c-T_0}$}
		\caption{\small Bubble action controlling the nucleation probability. The solid black curve corresponds to the microscopic result provided by holography. The blue, dot-dashed curve corresponds to the numerical fit in Eq.~\eqref{eq:fit_T0}, which is compatible with $\Delta F \propto(T-T_0)^{3/4}$. The dashed orange curve represents the thin-wall approximation, Eq.~\eqref{eq:thin_wall}, with a surface tension given by Eq.~\eqref{tension}. }
		\label{fig:S3overT_holographic}
	\end{figure}

	A homogeneous, metastable state decays to the stable, thermodynamically preferred state via the nucleation of critical bubbles. In the semiclassical approximation, the probability to nucleate a bubble per unit volume and unit time -the nucleation rate- is given by (see e.g.~\cite{Laine:2016hma})
	\begin{equation}\label{eq:probability}
		{\cal P}(T) = 
		a_0 \,e^{-\Delta F/T}\,,
	\end{equation}
	where the prefactor $a_0$ is  usually of order $a_0 \sim T^4$. We see that the probability is exponentially sensitive to the free energy difference. We will refer to the quantity $\Delta F/T$ as ``the bubble action''.  This quantity is given by Eq.~\eqn{eq:free_energy2} and is plotted in Fig.~\ref{fig:S3overT_holographic}. Let us now analyze the two limiting regimes, $T\to T_c$ and $T\to T_0$, which are clearly visible in the figure.
	
	Close to the critical temperature $\lw\ll R$ and the \textit{thin wall approximation} applies \cite{Laine:2016hma}. As mentioned in Sec.~\ref{sec:Critical_Bubbles}, in this regime the profile of a bubble is well  approximated by a hyperbolic tangent---see Eq.~\eqref{eq:radius_width1}--- and $\Delta F$ decomposes into two contributions: one from the  pressure difference (times the volume of the bubble) and a second one from the wall itself, proportional to the area of the bubble:
	\begin{equation}\label{eq:thin_wall_limit}
		\Delta F \simeq -\frac{4\pi \radius^3}{3}\Delta p + 4\pi \radius^2 \st\,.
	\end{equation}
	Here,  $\Delta p = p_{-} - p_+ > 0$ is the pressure difference between the low-energy stable and the high-energy metastable homogeneous phases, $\radius$ is the radius of the bubble, and $\st$ is the \textit{surface tension}. This expression is  motivated by the observation that the thickness of the bubble saturates as $T_c$ is approached---see Fig.~\ref{fig:radius_and_wall}. Extremizing Eq.~\eqref{eq:thin_wall_limit} we obtain an expression for the radius of the critical bubble in this regime:
	\begin{equation}
		\label{eq:thin_wall_radius}
		\radius \simeq \frac{2\st}{\Delta p} \,.
	\end{equation}
	Note that both $\st$ and $\Delta p$ scale as $a \sim N^2$, so $R$ is $N$-independent. 
	Substituting \eqref{eq:thin_wall_radius} into \eqref{eq:thin_wall_limit} we find 
	\begin{equation}
		\label{eq:thin_wall}
		\Delta F \simeq \frac{16\pi}{3}\frac{\st^3}{( \Delta p )^2}\,.
	\end{equation}
	The pressure difference vanishes at $T=T_c$, so near $T_c$ we have 
	\begin{equation}
		\Delta p \propto (T-T_c) \qquad 
		\Rightarrow \qquad
		\Delta F \propto (T-T_c)^{-2} \,.
	\end{equation}
	This  is confirmed by Fig.~\ref{fig:S3overT_holographic}, where this behavior is indicated by the fitted dashed orange curve. 
	
	From the fit in the figure we also obtain the surface tension 
	\begin{equation}
		\label{tension}
		\st \,\,\, \simeq \,\,\, 9.64\times 10^{-3}\, a\fs^3 
		\,\,\, \simeq \,\,\, 
		1.55 \times 10^{-1} \, a T_c^3 \,.
	\end{equation}
	As expected from large-\(N\) counting arguments, this quantity scales as \(a \propto N^2\). However, its  value is numerically suppressed relative to the naive dimensional estimates \(\st \sim a \fs^3\) and \(\st \sim a T_c^3\), respectively. Note that even a mild suppression in $\st$ can lead to a significant suppression of the bubble action. Interestingly, lattice studies~\cite{Lucini:2005vg,Salami:2025iqq} have found
	analogous behavior in the deconfinement transition of large-\(N\) \(SU(N)\) YM theories, leading to significantly less supercooling than expected from the na\"{\i}ve scaling~\cite{Agrawal:2025xul}. We will show that the same phenomenon arises in our setup and that, within the effective description of the bubbles, it can be traced back to a suppression of the kinetic term.
	
	In the opposite limit, close to $T_0$, we find 
	\begin{equation}\label{eq:fit_T0}
		\Delta F \propto (T-T_0)^{x}\,,
	\end{equation}
	with $x\simeq0.76$. This result is obtained from a power-law fit to the ten data points closest to $T_0$, which span one order of magnitude: 
	\begin{equation}
		1.2\times 10^{-4}\, \leq\, \frac{T-T_0}{T_c-T_0} \, \leq \, 1.3\times 10^{-3}\,.
	\end{equation}
	We believe that our result is compatible with $x = 3/4$ within our numerical precision. Interestingly, this is different from the exponent $x = 3/2$ found in  Ref.~\cite{Enqvist:1991xw}. We will come back to this point below.

	\section{Effective description}
	\label{sec:effective}
	The holographic results above provide a fully microscopic description of the critical bubbles. In this section we will use them to verify what features are well captured by a local, two-derivative effective action, which is frequently used in the literature. This effective action takes the form
	\begin{equation}
		\label{eq:effective_action}
		S_{\text{\tiny eff}}(T) =\int_0^{1/T}\dd\tau\int\dd^ 3x \parent{\frac{1}{2} {Z(\field,T)} (\nabla\field)^2 +\Veff(\field,T)}\,,
	\end{equation}
	where $\field$ is some order parameter that jumps discontinuously across the phase transition. Note that we are omitting a  $(\partial_\tau\field)^2$ term, since we are only considering O(3)-symmetric configurations. The choice of order parameter is not unique, and a natural choice for $\field$ is the VEV of the scalar operator dual to the bulk field $\phi$, namely 
	\begin{equation}\label{eq:oreder_parameter_Lambda}
		\field = \VEV\,.
	\end{equation}
	Note that this field has dimensions of (energy)$^{3}$, and consequently $Z(\varphi,T)$ has dimensions of (energy)$^{-4}$. In addition, since the effective action and $\field$ scale as $N^2$, it follows that  $Z\sim N^{-2}$. 
	
	The O(3)-symmetric critical bubbles are  classical solutions of the  equations of motion that follow from $S_{\text{\tiny eff}}$. In the spherically symmetric case they  reduce to one equation   
	in terms of the radial coordinate $\rho$:
	\begin{equation}\label{eq:eom_effective_bubble}
		\frac{\dd^2 \field}{\dd \rho^2} + \frac{2}{\rho} \frac{\dd \field}{\dd \rho} + \frac{1}{2} \frac{\partial_\field Z(\field,T)}{Z(\field,T)} \left( \frac{\dd \field}{\dd \rho} \right)^2 - \frac{\partial_\field \Veff(\field,T)}{Z(\field,T)} = 0\,,
	\end{equation}
	subject to the boundary conditions $\field'(0)=0$ and $\field(\infty) = \field_+$, with $\field_+$ the value of the order parameter in the metastable state with temperature $T$.
	
	The effective action is completely determined by the microscopic theory, in which it can be computed via a Legendre transform of the generating functional. We will review this procedure in Section~\ref{sec:Legendre}, in Section~\ref{holoderi} we will use it to compute the  effective potential in our holographic model following Ref.~\cite{Ares:2021ntv}, and in Section~\ref{sec:kinetic} we will obtain the kinetic term and discuss some issues related to scheme dependence. We will then find the critical bubbles in this effective description an compare them to those in the microscopic description. In phenomenological analysis some simplifying assumptions about the functional form of $Z$ and $\Veff$ are often made. In  Section~\ref{sec:polynomial} we will assess the validity of these approximations in our case. 
	
	\subsection{Definition of the effective action}
	\label{sec:Legendre}
	Here we will review the construction of the effective action  following the presentation in  Ref.~\cite{Ares:2021ntv}. The construction is based on the quantum effective action evaluated around homogeneous equilibrium states.
	
	The quantum effective action $\Gamma[\field]$, also known as the one-particle-irreducible (1PI) effective action, is defined as the Legendre transform of the Euclidean generating functional
	\begin{equation}
		W[J] = -\log Z[J] \, ,
	\end{equation}
	where $Z[J]$ denotes the path integral in the presence of an external source $J(x)$. Explicitly, the effective action is given by
	\begin{equation}\label{eq:definition_effective_action}
		\Gamma[\field] \equiv W[J] - \int \dd^4 x \, J(x)\,\field (x)\,,
	\end{equation}
	with the Euclidean spacetime integral defined as
	\[\int \dd^4 x \equiv \int_0^{1/T} \dd\tau \int \dd^3 x \equiv  V_3/T\, . \]
	In this expression, $\field(x)$ is the expectation value of the operator sourced by $J(x)$, given by the variation of the generating potential with respect to the source,
	\begin{equation}
		\label{WW}
		\frac{\delta W[J]}{\delta J(x)} = \field(x)\,.
	\end{equation}
	Similarly, by construction, the source can be obtained as the variation of the effective action with respect to $\field$: 
	\begin{equation}
		\label{eq:def_potential}
		\frac{\delta\Gamma[\field]}{\delta\field(x)} = -J(x)\,.
	\end{equation}
	This  equation has an important consequence, as it implies that any sourceless state of the theory can be obtained by solving the classical equations of motion derived from the effective action. Unfortunately, finding $\Gamma[\field]$ exactly is in general not possible, but progress can still be made with some simplifying assumptions. 
	
	Let us first assume that, in the absence of sources, the partition function is dominated by a saddle point with a given value of the expectation value $\field(x)=\field_0(x)$. Then, the generating functional can be expanded in the source as 
	\begin{equation}\label{eq:generatingfunc}
		W[J]=W_0+\int \dd^4x J(x)\field_0(x)+\frac{1}{2}\int \dd^4x \dd^4 y\, G_0(x,y) J(x)J(y)+\cdots \,,
	\end{equation}
	with $G_0$ the connected correlator of the operator evaluated at the saddle point. The expectation value in the presence of a non-zero source becomes
	\begin{equation}
		\field(x)=\frac{\delta W[J]}{\delta J(x)}=\field_0(x)+\int \dd^4 y \, G_0(x,y) J(y)+\cdots \,.
	\end{equation}
	From this last expression we can solve for the source in terms of the expectation value using the inverse correlator
	\begin{equation}
		J(x)=\int \dd^4 y\, G^{-1}_0(x,y)
		\Big[ \field(y)-\field_0(y) \Big]+\cdots \,.
	\end{equation}
	Together with Eq.~\eqref{eq:generatingfunc}, this can be used in \eqref{eq:definition_effective_action} to get an expression for the effective action to leading order:
	\begin{equation}\label{eq:effective_action_expansion}
		\Gamma[\field]=W_0-\frac{1}{2}\int \dd^4 x \dd^4 y\, G_0^{-1}(x,y) 
		\Big[ \field(x)-\field_0(x)\Big] \Big[ \field(y)-\field_0(y) \Big]+\cdots\,.
	\end{equation}
	We emphasize that this expression is an expansion about the saddle $\field_{0}(x)$ and no gradient expansion has been made at this stage. Neglecting the higher order terms represented by~``$\cdots$'', the equations of motion derived from Eq.~\eqref{eq:effective_action_expansion} are
	\begin{equation}\label{eq:eomsGamma}
		\frac{\delta \Gamma[\field]}{\delta \field(x)}=-\int \dd^4 y \, G_0^{-1}(x,y)\Big[ \field(y)-\field_0(y) \Big] = 0\,.
	\end{equation}
	By construction, the solutions correspond to configurations with zero source (see Eq.~\eqref{eq:def_potential}), which can be expressed as
	\begin{eqnarray}\label{eq:small_perturbation_kernel}
		\field(x)=\field_0(x)+\eta(x)\,,
	\end{eqnarray}
	where $\eta(x)$ belongs to the kernel of the inverse propagator.
	
	We will now apply this general discussion   to the construction of the effective action $S_{\text{\tiny eff}}(T)$ in Eq.~\eqref{eq:effective_action}. For this purpose, we begin by choosing a homogeneous equilibrium state $\field_\ho$ as the state around which we expand, i.e., we set $\field_0(x)=\field_\ho$, with $ \partial_\mu \field_\ho=0$. Since the quantum effective action coincides with the free energy, for a homogeneous state we find that 
	\begin{equation}
		\Gamma[\field_\ho]=W_0[\field_\ho]= \frac{F_\ho}{T} =  \frac{V_3}{T}\, \Veff(\field_\ho,T)\,.
	\end{equation}
	Moreover, the requirement that $\field_\ho$ be an extremum of the effective action reduces, for a homogeneous configuration, to the condition that it extremizes the effective potential, $\partial_\field \Veff(\field_\ho,T)=0$.
	
	To get an expression for the kinetic term we need to consider inhomogeneous states. 
	Let us linearize the effective action $S_{\text{\tiny eff}}(T)$ from Eq.~\eqref{eq:effective_action} around the equilibrium state $\field_\ho$, in the way suggested by Eq.~\eqref{eq:small_perturbation_kernel}. In Fourier space, indicated by quantities with tildes, the linearized equations of motion read
	\begin{equation}
		\label{eq:eomsfluct}
		\left[ Z(\field_\ho,T) q^2 +\partial_\field^2 \Veff(\varphi_\ho,T)\right] \tilde \eta(\omega=0,q)=0 \,.
	\end{equation}
	
	We impose that the solutions to these equations coincide with the solutions obtained from the quantum effective action,  Eq.~\eqref{eq:eomsGamma}. It follows that the Fourier transform of the inverse correlator, $\widetilde G_0^{-1}$, has a zero at values of the spatial momentum modulus $q^2 = P_0$ for which Eq.~\eqref{eq:eomsGamma} vanishes. Close to these values
	\begin{equation}\label{eq:inverseGreenfunc}
		- \widetilde G_0^{-1}(\omega=0,q^2\to P_0)
		\,\,\sim\,  Z(\field_\ho,T) q^2+\partial_\field^2 \Veff(\varphi_\ho,T) \,\,\to\,\, 0\,.
	\end{equation}
	We have fixed a possible relative factor by demanding $\Gamma[\field]=S_{\text{\tiny eff}}(T)$ at quadratic order. The same value of $q^2$ corresponds to a pole of the correlator $\widetilde G_0$. Sufficiently close to this point
	\begin{equation}\label{eq:pole}
		\widetilde G_0(\omega=0,q^2\to P_0)\sim \frac{R_0}{q^2-P_0}\,.
	\end{equation}
	Inverting this expression, we see that the location of the pole, $P_0$, and the value of the residue, $R_0$, are related to the potential and kinetic terms of the effective action through
	\begin{equation}\label{eq:polevalues}
		P_0=-\frac{\partial_\field^2 \Veff(\varphi_\ho,T)}{Z(\field_\ho,T)},\quad R_0=-\frac{1}{Z(\field_\ho,T)}\,.
	\end{equation}
	If the location and the residue of the pole are known, these relations can be used to extract the coefficient of the kinetic term, $Z(\field_\ho,T)$. In principle there can be more than one pole contributing to the two-point function. We discuss this possibility in more detail in Appendix~\ref{app:effact}. Note also that, strictly speaking, this only fixes the effective action up to quadratic order in the expansion around the equilibrium configuration. In order to fully determine the effective action we would need to compute higher order correlation functions that would enter in the expansion of the quantum effective action \eqref{eq:effective_action_expansion} and introduce non-linear terms in the equations of motion \eqref{eq:eomsGamma}. Since this is not feasible in practice, we will take the quantum effective action evaluated on homogeneous equilibrium states as a proxy for the resummation of all non-linear contributions.
	
	\subsection{Holographic construction of the effective potential}
	\label{holoderi}
	In this section we will use holography to derive the effective potential for the order parameter from first principles. We will then compare the results from this effective description to the microscopic ones. 
	
	To connect the previous discussion to our holographic model, let us consider a fixed temperature $T$ and a fixed coupling $\fs$. This selects gravity solutions with a fixed value of the ratio $\fs/T$ as the duals of saddle points of the QFT for which the source vanishes, $J(x)=0$. The free energy of these saddle points can be found by drawing a vertical line in Fig.~\ref{fig:free_energy} at the corresponding value of $\fs/T$. Changing the temperature modifies the value of $\fs/T$ that defines the sourceless saddle points, causing the vertical line to shift horizontally in the figure. Once these reference points are identified, we can consider solutions with a non-zero constant source $J(x)=J$, which will be those for which the leading order coefficient of the scalar field, see Eq.~\eqref{UVexp_alt}, is the sum of the coupling and the source, $\source=\fs + J$.
	
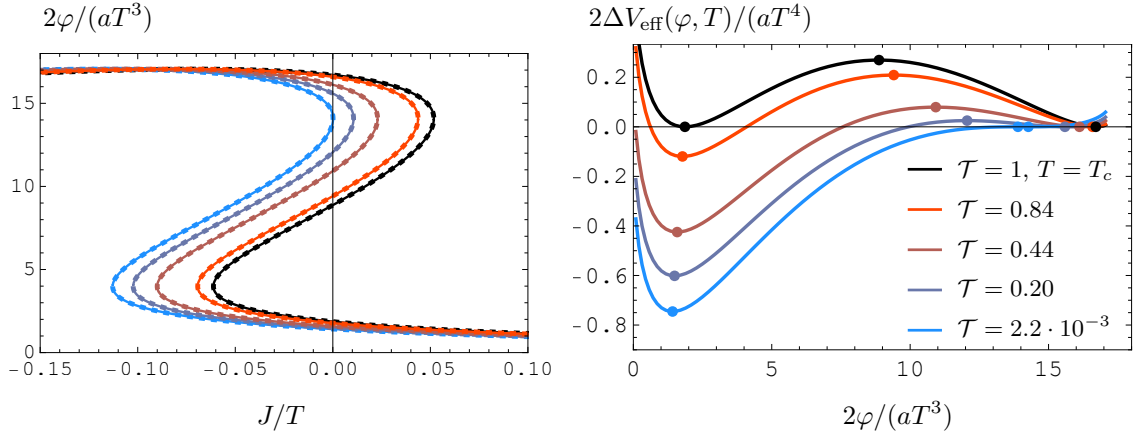
\begin{figure}[t]
		\centering
		\includegraphics[width=0.49\linewidth]{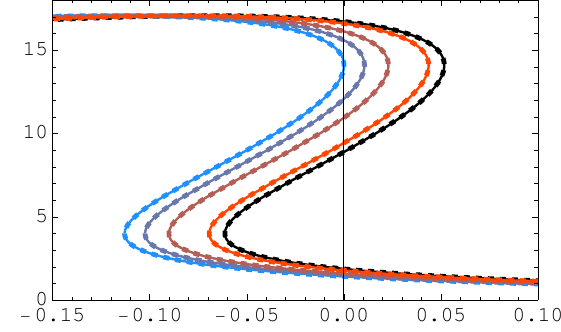}
		\put(-190,135){\small $\displaystyle 2\field /(aT^3)$}
		\put(-110,-10){\small $J/T$}
		\hfill
		\includegraphics[width=0.49\linewidth]{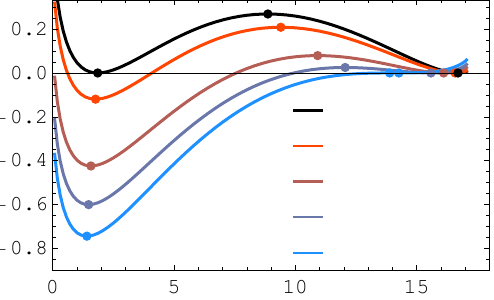}
		\put(-205,135){\small $2 \Delta\Veff(\field,T)/(aT^4)$}
		\put(-110,-10){\small $ \displaystyle 2\field /(aT^3)$}
		\put(-67,78){\footnotesize $\ratioT = 1$, $T=T_c$}
		\put(-67,63){\footnotesize $\ratioT = 0.84$}
		\put(-67,48){\footnotesize $\ratioT = 0.44$}
		\put(-67,33){\footnotesize $\ratioT = 0.20$}
		\put(-67,18){\footnotesize $\ratioT = 2.2\cdot 10^{-3}$}
		\caption{\small (Left) Order parameter  $\field$ as a function of the source $J$. Solid and dashed   curves correspond to Eqs.~\eqref{dowe} and~\eqref{eq:operator_from_derivative}, respectively. (Right) Effective potential obtained from the Legendre transform. The chosen values of the relative temperature difference $\ratioT$, defined in Eq.~\eqref{eq:rel_temperature_difference}, are shown in the legend. Here, $\be = \lambda_4 = -1/4$, which corresponds to the choice for which the ground state has zero energy.}
		\label{fig:operator_effective}
	\end{figure}

	From the gravity dual, Eq.~\eqref{eq:formula_VEV} gives the expectation value of the dual operator as a function of $\source$, which, for a homogeneous configuration, reduces to
	\begin{equation}
		\label{dowe}
		\field=  \VEV = 
		\frac{2L^3}{\kappa_5^2}\parent{{-2\fv(\phi_s) -4\be \phi_s^ 3}}\,.
	\end{equation}
	For fixed values of $\fs$ and $T$, this expression  determines the expectation value of the operator as a function of the source:  $\field=\field(J)$. The same function can be obtained from the derivative of the free energy density with respect to the source at fixed $\fs$ and $T$ (cf.~Eq.~\eqn{WW}):
	\begin{equation}\label{eq:operator_from_derivative}
		\field (J)= \frac{\partial f}{\partial J}\bigg|_{T,\fs}\,.
	\end{equation}
	For each choice of $\fs/T$ we obtain a different curve for $\field$ as a function of $J$, as shown in Fig.~\ref{fig:operator_effective}(left).\footnote{It is insightful to note that the curves in Fig.~\ref{fig:operator_effective} (left) are shifted horizontally. This is because the source $J$ (here constant) appears in the action exactly the same way as the coupling~$\fs$; since different curves correspond to different~$\fs$ (at fixed $T$) it is natural that the curves are shifted in $J$.} 
	We can invert this relation in the region where it is injective to obtain the source as a function of the expectation value,  $J=J(\field)$, and the effective potential is obtained by integrating this function at fixed $\fs$ and $T$ (see Eq.~\eqn{eq:def_potential}):
	\begin{equation}
		\label{eq:effective_potential_integral_source}
		{\Veff}(\field,T)={\Gamma}[{\field}]=f(\fs/T)-\int_{{\varphi}_0}^{{\field}} \dd{\field}' J({\field}')\,, \qquad \varphi_0=\varphi(J=0)\,. 
	\end{equation}
	The result is shown in Fig.~\ref{fig:operator_effective}(right) in terms of the difference between the potential and its value in the metastable state, 
	\begin{equation}\label{eq:deltaV}
		\Delta \Veff(\field,T) = 
		\Veff(\field,T) -  \Veff(\field_+,T) \,.    
	\end{equation}
	In the figure we see the existence of two minima and one maximum in the region of the phase transition, as expected.\footnote{A similar result for the same model was obtained in \cite{jorge}.} With the effective potential in hand, we can proceed to the determination of the kinetic-term coefficient $Z(\field,T)$.
	
	Before doing so, note that the quantities shown in Fig.~\ref{fig:operator_effective} are scheme dependent, due to the explicit appearance of $\be$ in the formulas for the expectation value of the dual operator and the free energy density. In the calculation of the kinetic term we encounter an additional issue related to scheme dependence. We will explain this in more detail below.
	
	\subsection{Scheme dependence and kinetic term}\label{sec:kinetic}
	
	The scheme dependence arises if one follows the procedure of Refs.~\cite{Ares:2021ntv,Henriksson:2025vci}, but was absent in those works because the scaling dimension of the scalar operator considered there forbids the introduction of finite counterterms. The strategy adopted in those references was to compute the two-point function $\tilde G_0$ at zero frequency to quadratic order in an expansion in $q^2/T^2$, then take the inverse to obtain $\tilde G_0^{-1}$ to the same order, and finally identify $Z(\field,T)$ by comparing the result with Eq.~\eqref{eq:inverseGreenfunc}.  However, when the theory allows for finite counterterms, the terms in the low momentum expansion of the inverse correlator become scheme dependent due to the possible presence of contact terms. In the case at hand there are two possible contact terms appearing in the correlator
	\begin{equation}
		\widetilde G(\omega=0,q)=\frac{R_0}{q^2-P_0}+c_0 +c_2 q^2 
	\end{equation}
	Then, the inverse of the correlator expanded at low momentum becomes
	\begin{equation}\label{eq:pollutedinvG}
		\widetilde G^{-1}(\omega=0,q)=-\frac{P_0}{R_0-c_0 P_0}-\frac{R_0-c_2 P_0^2}{(R_0-c_0  P_0)^2}q^2+\cdots.
	\end{equation}
	We see explicitly that the low momentum expansion gets polluted by the contact terms.
	
	Applying the low momentum expansion to our case, the result for the inverse correlator reads (see Appendix~\ref{app:greenfunc})
	\begin{equation}
		\label{eq:small_expansion_Green_function}
		\widetilde G_0^{-1}(\omega=0,q) = A_0(\be)+B_0(\alpha,\be) q^2+O(q^4),
	\end{equation}
	where the precise expressions for the coefficients $A_0(\be)$ and $B_0(\alpha,\be)$ are given in Eq.~\eqref{eq:Zalpha_dependent}. 
	Together with $\be$, here we observe the appearance of $\alpha$, an additional  arbitrary  coefficient of a local counterterm, see Eq.~\eqref{eq:ct_app}.
	Inverting Eq.~\eqref{eq:small_expansion_Green_function} and comparing the result to Eq.~\eqref{eq:inverseGreenfunc} 
	na\"{\i}vely suggests that we should identify 
	\begin{equation}
		\label{eq:kinetictermbad}
		\partial_\varphi^2 \Veff(\varphi,T)= -A_0(\be) \,,\qquad Z(\varphi,T) = - B_0(\alpha,\be)\,.
	\end{equation}
	
	\begin{figure}
		\centering
		\begin{center}
			\underline{$\be = \lambda_4 = -1/4$}
		\end{center}
		\includegraphics[width=0.48\linewidth]{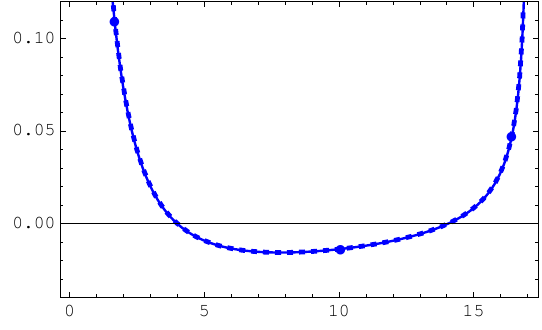}
		\put(-200,130){$\partial_\field^2\Veff(\field,T)\times aT^2/2$}
		\put(-120,-10){$ \displaystyle  2\field /(aT^3)$}
		\hfill
		\includegraphics[width=0.48\linewidth]{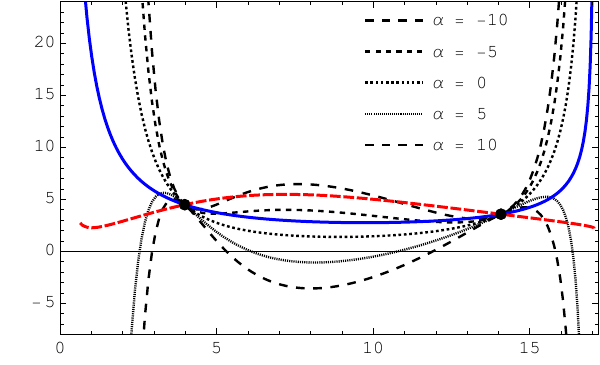}
		\put(-210,130){$Z(\field, T)\times 10^3aT^4$}
		\put(-120,-10){$ \displaystyle 2\field /(aT^3)$}
		
		\begin{center}
			\underline{$\be = 0$}
		\end{center}
		
		\includegraphics[width=0.48\linewidth]{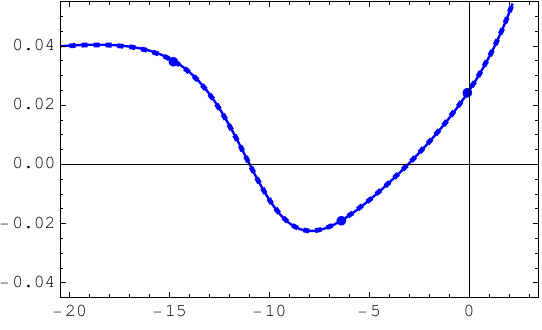}
		\put(-200,130){$\partial_\field^2\Veff(\field,T)\times aT^2/2$}
		\put(-120,-10){$ \displaystyle  2\field /(aT^3)$}
		\hfill
		\includegraphics[width=0.48\linewidth]{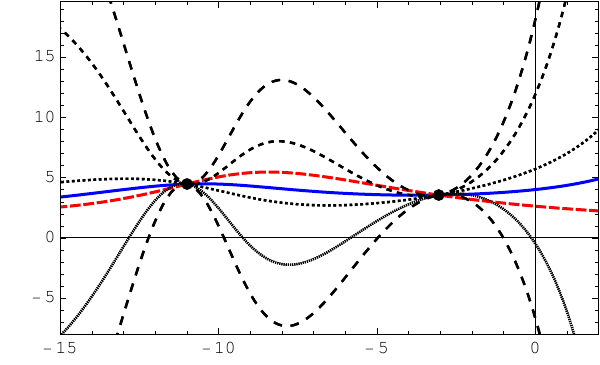}
		\put(-210,130){$Z(\field, T)\times 10^3aT^4$}
		\put(-120,-10){$ \displaystyle 2\field /(aT^3)$}
		\caption{\small 
			For the two different values of $\be$ indicated we show: (Left) Second derivative of the effective potential as a function of $\field$, obtained from Eq.~\eqref{eq:gamma2} and the first equation in \eqref{eq:kinetictermbad} (solid black), and from taking the second derivative of the potential reconstructed by integrating the source in Eq.~\eqref{eq:effective_potential_integral_source} (dashed blue). For illustration, the blue dots correspond to the equilibrium configurations at $T\simeq0.3932\fs$, corresponding to $\ratioT \simeq 0.66$, where $\ratioT$ is defined in Eq.~\eqref{eq:rel_temperature_difference}. (Right) The $\alpha$-independent coefficient of the  kinetic term,  obtained using the pole of the propagator as in Eq.~\eqref{eq:kinetic_from_Vpp} (see also Eq.~\eqref{eq:kinetic term_final}) is shown in solid blue. The black curves correspond to the scheme-dependent results that one would obtain by expanding the inverse propagator to $O(q^2)$. The dashed red curves stand for the value of minus the residue at the pole, which should coincide with the kinetic term according to Eq.~\eqref{eq:polevalues}. The black dots correspond to the value of the order parameter at the turning points, where the pole is at $q^2=0$ and the scheme dependence  disappears. 
		}
		\label{fig:kinetic}
	\end{figure} 
	
	The fact that Eqs.~\eqref{eq:small_expansion_Green_function} and~\eqref{eq:kinetictermbad} depend on arbitrary coefficients, as illustrated in Fig.~\ref{fig:kinetic}, is consistent with Eq.~\eqref{eq:pollutedinvG}, showing that the effective action constructed from the low momentum expansion is scheme-dependent and hence physically ambiguous. 
	
	The first term in \eqref{eq:small_expansion_Green_function} matches the second derivative of the effective potential obtained from Eq.~\eqref{eq:effective_potential_integral_source}, as illustrated  in Fig.~\ref{fig:kinetic} (left). Note that, although the effective potential depends non-trivially on the choice of $\fs/T$, its second derivative $\partial_\field^2\Veff(\field,T)$ does not, as illustrated by the figure. Despite its scheme dependence, the effective potential derived from the Legendre transform has the right expected properties, with extrema at the equilibrium values of the expectation value of the operator, where it correctly reproduces the corresponding free energy. Given that we do not know a full expansion of the effective action, as explained in subsection \ref{sec:Legendre}, this is actually our best candidate. The second term in \eqref{eq:small_expansion_Green_function}, on the other hand, would determine the coefficient of the kinetic term in the effective action, but it has no redeeming property that motivates completely overlooking its additional scheme dependence.
	\begin{figure}[t]
		\centering
		\includegraphics[width=1.05\linewidth]{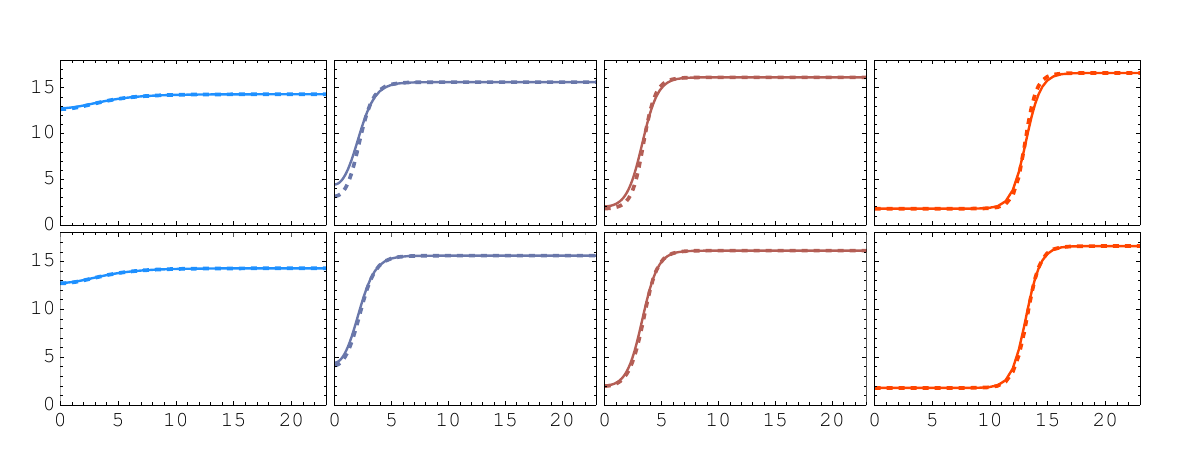}
		\put(-420,165){\small $2\overline \field/(aT^3)$}
		\put(-40,3){\small $\rho\fs$}
		\put(-140,3){\small $\rho\fs$}
		\put(-240,3){\small $\rho\fs$}
		\put(-340,3){\small $\rho\fs$}
		\put(-455,40){\rotatebox{90}{\small \underline{{$\be = 0$}}}}
		\put(-455,105){\rotatebox{90}{\small \underline{{$\be = \lambda_4$}}}}
		\caption{\small Comparison between the bubble profiles obtained from the microscopic holographic computation  (solid) and the ones obtained from the effective action (dashed) for $\be = \lambda_4 = -1/4$ (top) and $\be = 0$ (bottom). To meaningfully compare different choices of $\be$, we have defined shifted the field defining $\overline \field = \field - 4(\lambda_4-\be)\Lambda^3$.}
		\label{fig:bubbles_effective_holo}
	\end{figure}
	
	The dependence on $\alpha$ can be avoided if we identify the pole $q^2=P_0$ of the Green's function that is closer to $q^2=0$ and compute directly the kinetic term using Eq.~\eqref{eq:polevalues}, which gives
	\begin{equation}\label{eq:kinetic_from_Vpp}
		Z(\field_\ho,T) = - \frac{\partial_\field^2 \Veff(\varphi_\ho,T)}{P_0}\,.
	\end{equation}
	Defined in this way, the kinetic term does not exactly coincide with minus the inverse of the residue at the pole, as required by the second relation in Eq.~\eqref{eq:polevalues}, as shown in Fig.~\ref{fig:kinetic}. The discrepancy inherits the scheme dependence and remains mild in the spinodal region, where the effects of the inhomogeneity are more pronounced. Despite this mismatch, we will see that our prescription reproduces the bulk results remarkably well.
	
	The mode that gives rise to this pole can be understood as follows. In the spinodal region, the sound mode is unstable at low momentum and stable at high momentum. Its dispersion relation takes the form of a parabola, as shown  in e.g.~Fig.~3 of \cite{Bea:2021zol}. By continuity, in the spinodal region there is a  mode with 
	\begin{equation}
		\omega=0 \,,\qquad  q^2 > 0 \,.
	\end{equation}
	The value of $q^2$ approaches zero as the turning point at $T=T_0$ is approached along the spinodal branch. Past this point, once in the metastable branch, this mode moves into the complex $q$-plane since now $q^2<0$. However, it remains closer to the origin than the rest of the poles and its residue gives the correct value of $Z(\field_\ho, T)$ through Eq.~\eqref{eq:polevalues}. The result is shown in red in Fig.~\ref{fig:kinetic} (right), overlaid on the $\alpha$-dependent kinetic terms previously obtained from Eq.~\eqref{eq:kinetictermbad}. As expected, $Z \sim a^{-1} \sim N^{-2}$. We find that $Z$ exhibits a strong dependence on $\field$ in the locally stable branches, i.e.~outside the region delimited by the black dots, while it is approximately field-independent along the spinodal branch, corresponding to the region between these points. Moreover, in this region the magnitude of $Z$ is suppressed by roughly three orders of magnitude relative to the naive dimensional estimate, $a Z \sim T^{-4}$. We will return to these features below.
	\begin{figure}[t]
		\centering
		\includegraphics[width=0.485\linewidth]{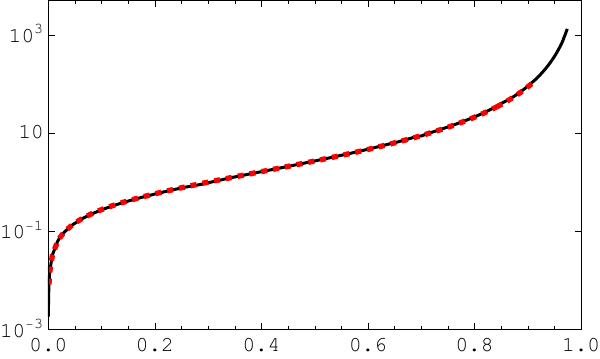}
		\put(-223,120){\small $\displaystyle \frac{2\Delta F}{a T}$}
		\put(-117,-17){\small $\displaystyle \frac{T-T_0}{T_c-T_0}$}
		\caption{\small Comparison of the bubble action, as obtained from the microscopic computation (solid black) and from an  effective action computed from first principles via holography (dashed red). In this plot we used $\beta = \lambda_4$.}
		\label{fig:bubbles_effective_holo_action}
	\end{figure}
	
	We now have all the ingredients required to determine the critical bubbles by solving Eq.~\eqref{eq:eom_effective_bubble} for different values of the temperature. The results for the bubble profiles are shown in Fig.~\ref{fig:bubbles_effective_holo} for different values of $\be$, where we see that the effective approach provides a remarkably accurate approximation to the exact result. The deviations become larger when the interior of the bubble corresponds to a state further away from the homogeneous equilibrium configurations, which is consistent with the fact that the effective action was derived by expanding around those states. Fig.~\ref{fig:bubbles_effective_holo_action} presents a similar comparison for the free energy excess, where once again a remarkable level of agreement is observed.

	\subsection{Phenomenological construction of the effective action}\label{sec:polynomial}
	We have seen that when the two-derivative effective action \eqref{eq:effective_action} is known, it can be used to make accurate predictions about bubble properties, particularly the bubble action. However, in many cases of interest the effective action cannot be determined from first principles. In some instances, this is because the microscopic theory is unknown, as in Beyond the Standard Model scenarios. In other cases, the microscopic theory is known, but calculating the effective action is not feasible due to technical challenges, such as the strongly coupled nature of the physics. Despite these obstacles, progress can still be made by constructing an effective action motivated by a combination of phenomenological considerations and dimensional analysis. In this section, we will evaluate the accuracy of this approach in our case by comparing it to the microscopic description.
	
	We begin with  the effective potential. Renormalizability, together with the requirement to accommodate a FOPT, leads to an ansatz in the form of a degree-four polynomial---see e.g.~\cite{Enqvist:1991xw,Cutting:2020nla,Hindmarsh:2020hop}. For example, in the context of EW-like phase transitions, the order parameter is the expectation value of the Higgs field,  $\langle\Phi_{\text{\tiny h}}\rangle$, and the potential is an expansion in powers of $\langle\Phi_{\text{\tiny h}}\rangle/T$ truncated at fourth order. 
	
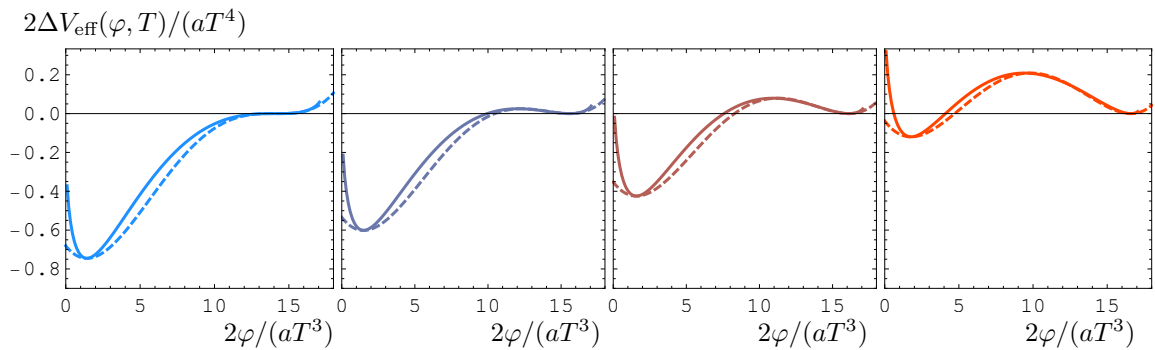
\begin{figure}[t]
		\centering
		\includegraphics[width=1.05\linewidth]{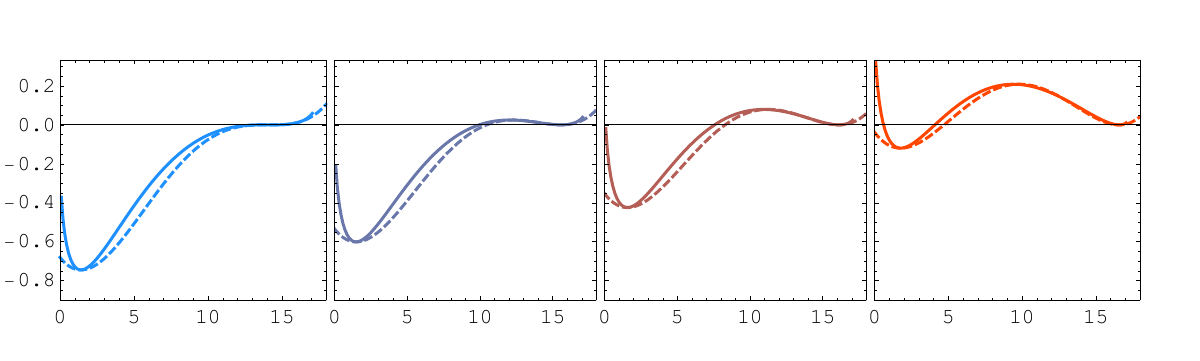}
		\put(-445,120){\small $2 \Delta\Veff(\field,T)/(aT^4)$}
		\put(-70,3){\small $2\field/(aT^3)$}
		\put(-170,3){\small $2\field/(aT^3)$}
		\put(-270,3){\small $2\field/(aT^3)$}
		\put(-370,3){\small $2\field/(aT^3)$}
		\caption{\small Comparison of the effective potentials shown in Fig.~\ref{fig:operator_effective} (right) (solid curves), obtained directly from the quantum effective action, with their approximations using a degree-four polynomial (dashed curves).}
		\label{fig:potentials_comparison}
	\end{figure}

	The requirement that the effective potential be consistent with the equation of state fixes the quartic polynomial. The expectation value of the order parameter, $\field=\VEV$, is a multivalued function of the temperature across the phase transition. In particular, for temperatures in the range $T\in(T_0,\Tc)$ it can take three distinct values, $\field_+$, $\field_s$, and $\field_-$, corresponding to the metastable ($+$), spinodal ($s$), and stable ($-$) branches, respectively. Locally stable and unstable phases are associated with minima and maxima of the effective potential. Consequently, throughout the region of the FOPT, the potential must exhibit two minima separated by a single maximum. Moreover, the value of the potential at each of these extrema must coincide with the corresponding free energy density. We will therefore impose
	\begin{equation}
		\label{eq:condition_polynomial_4}
		\begin{array}{ll}
			\partial_\field\Veff(\field,T)\big|_{\field = \field_+} = 0\,,\qquad&\qquad \Veff(T,\field_+) = f_+\,,\\[\spacelist]
			\partial_\field\Veff(\field,T)\big|_{\field = \field_-} = 0\,,    & \qquad \Veff(T,\field_-) = f_-\,,\\[\spacelist]
			\partial_\field\Veff(\field,T)\big|_{\field = \field_{\text{\tiny M}}} = 0 \,,\qquad&\qquad
			\Veff(T,\field_{\text{\tiny M}}) = f_s\,.
		\end{array}
	\end{equation}
	Note that we enforce the two minima of the potential to occur at the values of the order parameter $\field_+$ and $\field_-$ determined by the equation
	of state. By contrast, for the maximum we only require that the value of the potential reproduces the free energy density of the spinodal branch, $f_s$,
	without fixing its location, namely, we do not impose $\field_{\text{\tiny M}} = \field_s$. Enforcing this additional condition would overconstrain the potential, since a quartic polynomial has only five independent parameters. This asymmetry is motivated by the fact that  the locally stable states associated with the minima appear as asymptotic configurations of planar or large spherical bubbles, whereas the unstable state corresponding to the maximum does not. In Fig.~\ref{fig:potentials_comparison}, we compare the resulting degree-four polynomial potentials with those obtained from the quantum effective action, finding good agreement both qualitatively and quantitatively.
	
	We now turn to the coefficient of the kinetic term in the effective action, $Z(\field, T)$. Unlike the effective potential, this cannot be determined from the equation of state. For this reason, simplifying assumptions are often made. The first one is that $Z$ is approximately field-independent, i.e.~that $Z(\field,T) \approx Z(T)$. In the spherically-symmetric case, we can then rescale the radial coordinate through 
	\begin{equation}
		\rho = Z(T)^{1/2} \nrho
	\end{equation}
	and rewrite the effective action \eqref{eq:effective_action} as 
	\begin{equation}
		\label{eq:rescaled_effecttive_action}
		S_{\text{\tiny eff}}(T) =\frac{4\pi}{T}{Z(T)}^{3/2}\int_0^\infty\dd\nrho \nrho^2 \parent{\frac{1}{2}(\partial_\nrho\field)^2 +\Veff(\field,T)}\,.
	\end{equation}
	Note that $\nrho$ has dimensions of (energy)$^1$. This assumption simplifies the problem significantly because $Z$ only appears as an overall multiplicative factor in the action. Consequently, the equation of motion becomes independent of $Z$: 
	\begin{equation}\label{eq:eom_effective}
		\frac{\dd^2 \field}{\dd \nrho^2} + \frac{2}{\nrho} \frac{\dd \field}{\dd \nrho} - {\partial_\field \Veff(\field,T)} = 0.
	\end{equation}
	The net result is that we can determine the shape of the bubble profiles in the $\nrho$-coordinate without needing to know the normalization of the kinetic term. However, to obtain the overall physical size of the bubbles, i.e.~their size in  the $\rho$-coordinate, and to calculate the bubble action, which depends on the overall magnitude of $S_{\text{\tiny eff}}(T)$, the value of $Z(T)$ is required. The key point is that if the value of $Z$ is determined from the microscopic theory, then excellent agreement with the microscopic results is found, thus justifying the assumption that $Z$ is approximately field-independent. On the other hand, if $Z$ can only be estimated using dimensional analysis, then a significant discrepancy arises.
	
\begin{figure}
		\centering
		\includegraphics[width=1.06
		\linewidth]{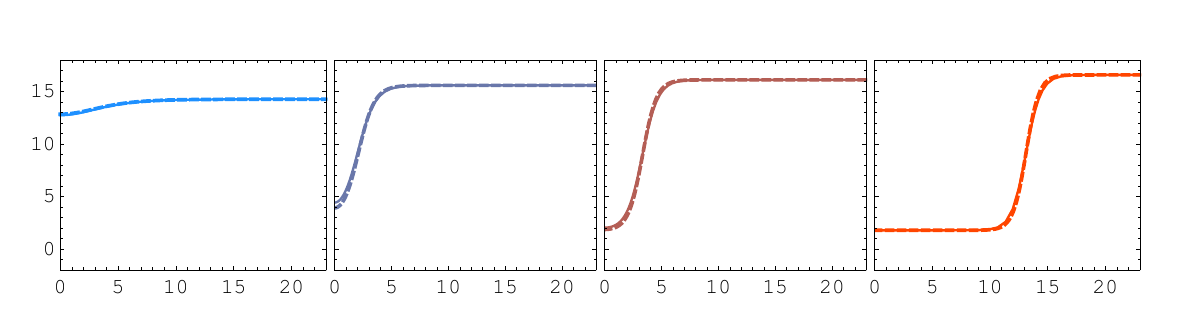}
		\put(-425,110){\small $2\field/(a T^3)$}
		\put(-40,3){\small $\rho\fs$}
		\put(-140,3){\small $\rho\fs$}
		\put(-240,3){\small $\rho\fs$}
		\put(-340,3){\small $\rho\fs$}
		\put(-427,28){\footnotesize $Z(T)aT^4\simeq 4.07\cdot10^{-3}$}
		\put(-325,28){\footnotesize $Z(T)aT^4\simeq 3.71\cdot10^{-3}$}
		\put(-222,28){\footnotesize $Z(T)aT^4\simeq 3.67\cdot10^{-3}$}
		\put(-120,28){\footnotesize $Z(T)aT^4\simeq 3.73\cdot10^{-3}$}
		\caption{\small Profiles of the bubbles as a function of the radial coordinate for the order parameter $\field = \VEV$. The four representatives shown correspond to the same temperatures as in the previous figures, see Fig.~\ref{fig:results_bubbles}.  The solid curves correspond to the result obtained from the full computation in the bulk, while the dashed ones are the result of the effective approach.}
		\label{fig:profiles_bubbles_from_pol}
	\end{figure}
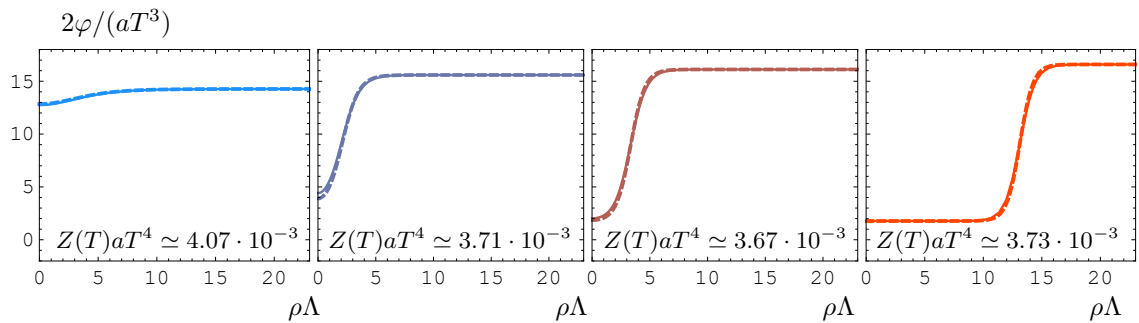

\begin{figure}[t]
		\vspace{2mm}    
		\hfill \includegraphics[width=0.45\linewidth]{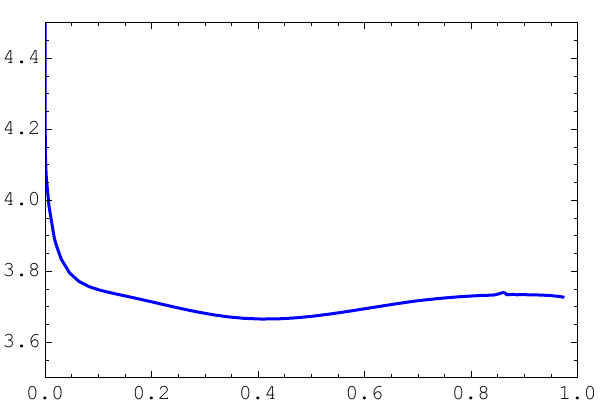}
		\put(-185,130){\small $Z(T)\times 10^{3}\,aT^4$}
		\put(-110,-15){\small $\displaystyle \frac{T-T_0}{T_c-T_0}$}
		\hfill
		\includegraphics[width=0.485\linewidth]{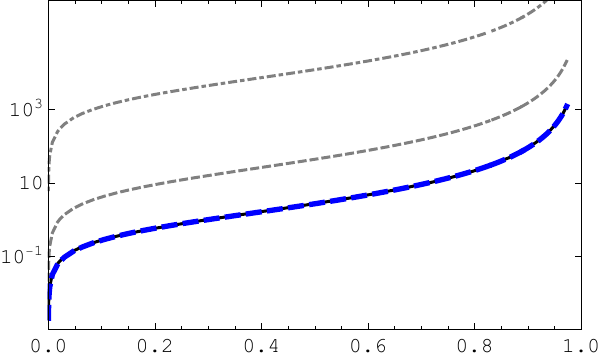}
		\put(-217,125){\small $\displaystyle \frac{2\Delta F}{a T}$}
		\put(-110,-15){\small $\displaystyle \frac{T-T_0}{T_c-T_0}$}
		\caption{\small 
			(Left) Coefficient of the kinetic term in the effective action with a degree-four polynomial that provides the best fit of the corresponding bubble profiles to those derived from the microscopic theory. (Right) Comparison between the bubble action obtained from the effective action with a degree-four polynomial (gray and blue curves) and the microscopic result (solid black curve). The dashed blue curve corresponds to the case where $Z(T)$ is fitted, as in the plot on the left. The gray dashed curve shows the result for the choice $a Z = \fs^{-4}$, while the dotted-dashed curve corresponds to $a Z= T^{-4}$.
		}
		\label{fig:VEV_and_effective_potential_4}
	\end{figure}

	To demonstrate this, we present the result of computing the bubble profiles in the $\nrho$-coordinate and then transforming them to the $\rho$-coordinate in Fig.~\ref{fig:profiles_bubbles_from_pol}. 
	For this transformation we use the value of $Z(T)$ that yields the best fit to the microscopic profiles. This fitted value is shown in Fig.~\ref{fig:VEV_and_effective_potential_4} (left), where we see that it is around
	\begin{equation}
		\label{actual}
		a Z \simeq 3.7
		\times 10^{-3} \, T^{-4} \,.
	\end{equation}
	This value agrees with that of $Z(\field, T)$   along the spinodal branch, corresponding to the region between the two black circles shown in Fig.~\ref{fig:kinetic} (right). The excellent agreement with the microscopic profiles seen in Fig.~\ref{fig:profiles_bubbles_from_pol} confirms that a field-independent $Z$ provides a good approximation, despite the fact that $Z$ depends strongly on $\field$  outside the spinodal branch. Presumably, the reason for this is that the kinetic term is only relevant in the region where gradients  are large, i.e.~across the bubble wall, and in this region the values of $\field$ correspond to those in the spinodal branch. Fig.~\ref{fig:VEV_and_effective_potential_4} shows that this procedure also yields an accurate prediction for the bubble action.  
	
	As mentioned earlier, in many relevant cases the coefficient $Z$ cannot be determined from first principles. In such situations, one can attempt to estimate its approximate value using dimensional analysis. In our case, two natural choices are:
	\begin{equation}
		a Z \,\,\, \sim \,\,\,  T^{-4}
	\end{equation}
	or
	\begin{equation}
		a Z \,\,\,\sim\,\,\, \Lambda^{-4} \,\,\,\sim \,\,\, 2.5  \times 10^{-1} \, T^{-4}.
	\end{equation}
	These estimates differ by 3 and 1 orders of magnitude, respectively, from the actual value derived from the microscopic theory---see Eq.~\eqref{actual}. Since the bubble action scales as $S_{\text{\tiny eff}} \sim Z^{3/2}$, this discrepancy leads to significant differences between predictions based on the dimensional estimate of $Z$ and those from the microscopic result, as shown in Fig.~\ref{fig:VEV_and_effective_potential_4} (right).
	
	The dimensional estimate can be substantially refined if additional information beyond the equation of state is available, specifically the surface tension. This is particularly useful given that the surface tension can be determined through lattice simulations as in Refs.~\cite{Lucini:2005vg, Salami:2025iqq}. Recall that we introduced the notion of surface tension in Eq.~\eqref{eq:thin_wall_limit} in the \textit{thin wall approximation}  (i.e. for large bubbles), where their structure can be decomposed into the excess free energy density (times the volume of the bubble) and a contribution from the wall:
	\begin{equation}
		\label{eq:rescaled_effecttive_action_Delta}
		\Delta S_{\text{\tiny eff}}(T)= \Delta S_{\text{\tiny inside}}(T) + S_{\text{\tiny wall}}(T)\,.
	\end{equation}
	In the effective theory, the first term is given by the difference in the effective potential, 
	\begin{equation}
		\Delta S_{\text{\tiny inside}}(T) = \frac{4\pi \radius^3}{3T}\Delta \Veff(\field_-,T)\,,
	\end{equation}
	with $\Delta \Veff (\field,T)$ defined by Eq.~\eqref{eq:deltaV}. Here, we will focus our attention  on the contribution from the wall. 
	
	In the large bubble regime in which the current analysis is applicable, the second term in the equation of motion \eqref{eq:eom_effective} is suppressed by the radius of the bubble, and thus negligible. Then
	\begin{equation}\label{eq:eom_thin_wall}
		\frac{\dd^2\field}{\dd\varrho^2} \simeq \partial_\field\Veff(\field,T)\quad \Rightarrow \quad \frac{1}{2}\left(\frac{\dd\field}{\dd \varrho} \right)^2 \simeq \Delta\Veff(\field,T)\,,
	\end{equation}
	where to get the second expression we have multiplied by $\partial_\nrho\field$ and integrated over $\varrho$ from infinity. Denoting the size of the bubble in the $\nrho$-coordinate by $\nrho_b \equiv R \, Z(T)^{-1/2}$, we can use Eq.~\eqref{eq:eom_thin_wall}---steps (\textit{i}) and (\textit{iii})---and the fact that $\nrho_b$ is almost constant in the region of the wall together with a change of variables---step (\textit{ii})---to evaluate 
	\begin{equation}
		\label{eq:rescaled_effecttive_action_wall}
		\begin{aligned}
			S_{\text{\tiny wall}}(T) &=\frac{4\pi}{T}{Z(T)}^{3/2}\int_{\nrho_b-\delta}^{\nrho_b+\delta}\dd\nrho \nrho^2 \parent{\frac{1}{2}(\partial_\nrho\field)^2 +\Delta \Veff(\field,T)}\\[\spacelist]
			&\overset{\text{\tiny (\textit{i})}}{\simeq} \frac{4\pi}{T}{Z(T)}^{3/2}\int_{\nrho_b-\delta}^{\nrho_b+\delta}\dd\nrho \ \nrho^2 {(\partial_\nrho\field)^2}\\[\spacelist]
			&\overset{\text{\tiny (\textit{ii})}}{\simeq}  \frac{4\pi}{T}{Z(T)}^{1/2}R^2\int_{\field_-}^{\field_+}\dd\field  \  {\partial_\nrho\field} \\[\spacelist]
			&\overset{\text{\tiny (\textit{iii})}}{\simeq}  \frac{4\pi}{T}{Z(T)}^{1/2}R^2\int_{\field_-}^{\field_+}\dd\field  \  \sqrt{2 \Delta\Veff(\field,T)} \\[\spacelist]
			& \equiv \frac{4\pi R^2}{T}\sigma\,.
		\end{aligned}
	\end{equation}
	The expressions above should be understood in the limit $\delta \to 0$ and $R \to \infty$. The latter ensures that the two minima, $\field_+$ and $\field_-$, of the effective potential become exactly degenerate, and that the potential difference $\Delta \Veff(\field,T)$, appearing under the square root, remains strictly positive between them. Thus, in the effective theory the surface tension can be expressed as
	\begin{equation}
		\label{integral_new}
		\sigma = Z(T_c)^{1/2}\int_{\field_-}^{\field_+} d\field \, \sqrt{2 \Delta\Veff(\field,T_c)} \,.
	\end{equation}
	At $T_c$ the two homogeneous solutions share the same value for the free energy density, 
	\begin{equation}
		f_+ = f_- \equiv f_c\,,    
	\end{equation}
	in which case the conditions in Eq.~\eqref{eq:condition_polynomial_4} imply that
	\begin{equation}
		\Delta\Veff(\field,T_c) = 16 (f_s - f_c) \frac{(\field-\field_+)^2(\field-\field_-)^2}{(\field_+ - \field_-)^4} \,.
	\end{equation}
	Substituting in \eqref{integral_new} we arrive at
	\begin{equation}\label{eq:surface_tension_pol_4}
		\sigma = Z(T_c)^{1/2} \, \frac{2\sqrt{2}}{3}\, |\Delta\field|\, \sqrt{f_s - f_c}\,.
	\end{equation}
	with $\Delta \field = \field_- -\field_+$. 
	In our case $\sigma$ is given in Eq.~\eqref{tension} and we can read off from Fig.~\ref{fig:operator_effective}(right) that
	\begin{equation}
		\Delta \Veff(\field_{\text{\tiny M}},T_c) = f_s-f_c\simeq 1.35 \times 10^{-1}\, aT_c^4 \,,\quad \qquad
		|\Delta \field| \simeq 7.42\, a T_c^3 \,.
	\end{equation}
	Substituting in \eqref{eq:surface_tension_pol_4}  and solving for $Z(T_c)$ gives 
	\begin{equation}
		a Z \simeq 3.66 \times 10^{-3} \, T_c^{-4} \,,
	\end{equation}
	which is remarkably close to the values that provide the best fit depicted in Fig.~\ref{fig:VEV_and_effective_potential_4}, see also Eq.~\eqref{actual}. Moreover, as in Eq.~\eqref{eq:fit_T0}, the scaling of the bubble action deviates from the standard result $\Delta F\propto(T-T_0)^{3/2}$ found in Ref.~\cite{Laine:2016hma}. The agreement between our microscopic result and the effective description based on a quartic polynomial with a constant kinetic term suggests that this discrepancy is not a consequence of the different approaches (microscopic vs effective), nor of the assumption that $Z(\field,T)$ is constant. Rather, we trace it to the way the effective potential develops an inflection point. Indeed, while in Ref.~\cite{Laine:2016hma} some coefficients of the effective potential vanish at the spinodal point, in our case all of them remain finite. It would be interesting to characterize the different exponents that one could get with a degree four polynomial, which we leave for future work.
	
	We conclude that requiring the effective action to reproduce the correct surface tension provides an accurate estimate of the kinetic term. As we have seen, under these conditions the effective action then reproduces the bubble properties with high precision.

	\section{Conclusions}\label{sec:conclusions}
	In this work we have presented a fully microscopic description of critical bubbles in a strongly coupled, four-dimensional gauge theory undergoing a first-order thermal phase transition. Using holography, we constructed static, inhomogeneous, and unstable black-brane solutions dual to $O(3)$-symmetric critical bubbles in the boundary theory. The explicit construction of these geometries allowed us to compute key properties of the critical bubbles directly from the microscopic theory across the entire metastable branch, including the bubble profile, the surface tension and the  nucleation rate. From the bulk perspective, these quantities are encoded in localized deformations of the black-brane horizon, providing a  geometric interpretation of the bubble properties.
	
	A central goal of this work was to assess the validity of effective descriptions of first-order phase transitions. To this end, we compared our microscopic results with those obtained from a two-derivative effective action for the order parameter in two distinct scenarios. When the effective action was derived holographically from first principles, we found remarkable agreement with the microscopic results for the bubble profiles, surface tension, and nucleation rates. This provides strong evidence that, when properly constructed, effective actions can faithfully capture even highly inhomogeneous and intrinsically nonperturbative configurations such as critical bubbles.
	
	In contrast, when the effective action is constrained solely by the equation of state and dimensional analysis, substantial discrepancies emerge. In particular, in our theory this approach significantly overestimates the coefficient of the kinetic term, leading to incorrect bubble properties and nucleation rates. We showed that these discrepancies can be traced to a suppression of the surface tension relative to naive dimensional expectations. Imposing the correct surface tension as an additional constraint on the effective action is sufficient to restore agreement with the microscopic results. Interestingly, lattice studies~\cite{Lucini:2005vg,Salami:2025iqq} have observed similar behavior in the deconfinement transition of large-$N$ $SU(N)$ Yang–Mills theories, implying markedly less supercooling than na\"{\i}vely expected~\cite{Agrawal:2025xul}. While this parallel is suggestive, it should be interpreted with caution, as in our case the transition takes place between two deconfined phases rather than between confined and deconfined phases.
	
	Several extensions of this work are natural. A particularly interesting one is the computation of the fluctuation determinant around the critical bubble, which controls the prefactor of the nucleation rate. Within the holographic framework, this problem can be naturally addressed by analyzing the spectrum of quasinormal modes of the corresponding inhomogeneous black-brane geometries \cite{Denef:2009kn}. Such a calculation would yield a fully microscopic determination of both the exponential suppression and the prefactor of the nucleation rate at strong coupling. 
	
	It would also be interesting to investigate whether a similar suppression of the surface tension occurs in other strongly coupled theories, and to what extent these insights can inform phenomenological models of cosmological or astrophysical first-order phase transitions. We expect that the framework developed here will provide a valuable benchmark for future studies of phase transitions beyond the reach of perturbative methods.

	\section*{Acknowledgements}
	We thank Alessio Caddeo, Jorge Casalderrey-Solana, Oliver Gould, Oscar Henriksson, Niko Jokela, Tomislav Prokopec, Ronnie Rodgers, Mikel Sanchez-Garitaonandia, Pedro Tarancon-Alvarez, Jorinde van de Vis and Miguel Vanvlasselaer for useful discussions.
	JS thanks Tomas Andrade, Alexander Krikun and Christiana Pantelidou for guidance in learning how to solve PDEs with pseudospectral methods.
	D.M. acknowledges financial support from Grant CEX2024-001451-M funded by MICIU-AEI-10.13039/501100011033, from Grant No. PID2022-136224NB-C22 from the Spanish Ministry of Science, Innovation and Universities, and from Grant No. 2021-SGR-872 funded by the Catalan Government.  C.H. is partially supported by the Spanish Agencia Estatal de Investigación and Ministerio de Ciencia, Innovacion y Universidades through the grants
	PID2021-123021NB-I00 and PID2024-161500NB-I00. This research is also funded by the European Union (ERC, HoloGW, Grant Agreement No. 101141909). 
	Views and opinions expressed are, however, those of the authors only and do not necessarily reflect those of the European Union or the European Research Council. Neither the European Union nor the granting authority can be held responsible for them.
	
	\appendix

	\section{Numerical Implementation} \label{app:implementation}
	In this Appendix we provide further details about how we solved the system of PDEs.
	
	To gain some handle on the numerical solutions and their precision, we performed some variable and coordinate redefinitions. First, note that in the UV expansion of the functions $Q_i(z,\rho)$ (see Eq.~\eqref{eq:UV_expansion_inhomogeneous}) the coefficients that have to be extracted from the numerical solution appear at relatively high order. In particular, we would need to take a fourth order derivative to compute them. Differentiating a numerical solution so many times introduce large numerical errors. For this reason, we preferred to solve the system of equations in terms of a new set of variables $q_i$, defined by
	\begin{equation}
		\begin{aligned}
			Q_i(\rho,z) &= 1 + z^3 q_i (\rho,z),\quad \text{for }i\in\{1, 2,3,4\}\\
			Q_5 (\rho,z) &= z^4q_5(\rho,z),\\
			Q_0 (\rho,z) &= 1+ z\,q_0(\rho,z)\,.
		\end{aligned}
	\end{equation}
	In terms of these, the UV boundary conditions are simply $q_i(z,x) = 0$, and the unknown coefficients are given in terms of first derivatives with respect to $z$. On the other hand, the new variables satisfy mixed boundary conditions at the horizon.
	
	In addition, it is also convenient to compactify the $\rho$ coordinate, for which we introduce a general coordinate change $\rho = \rho(x)$, with $x\in [0,1]$. The concrete form of $\rho(x)$ is adapted to the expected properties of each solution. For instance, for small bubbles (i.e. temperatures close to the spinodal point $T_0$), a convenient choice is
	\begin{equation}
		\rho(x) = \rho_1(x) \equiv \frac{\cnta \, x}{1-x^2}\, ,
	\end{equation}
	for some $\cnta\in\mathds{R}$. This coordinate is helpful for  bubbles for which the non-trivial, varying part of the solution is within a region close to the origin of a characteristic size~$\cnta$, as it happens for temperatures close to $T_0$. For each solution we adjust the value of the parameter $\cnta$.
	
	As the wall develops, we considered the more suitable yet involved coordinate change
	\begin{equation}
		\rho(x) = \rho_2(x) \equiv  \frac{\cnta\,  x \left(\cntd+3 \cntt^2-3 \cntt+1\right) \left(\cntd+3 \cntt^2+x^2-3 \cntt \, x\right)}{x^2 \left(\cntd+3 \cntt^2+x^2-3 \cntt \,x\right){}^2-\left(\cntd+3 \cntt^2-3 \cntt+1\right){}^2}\ .
	\end{equation}
	This coordinate and the corresponding choice of $\cnta$, $\cntd$, $\cntt$ $\in \mathds{R}$ are engineered so that the relevant region of the solution is zoomed in, at the same time that the wall is located around $x\simeq 1/2$. Note that $\rho_2 (x)= \rho_1(\cntd(x-\cntt) + (x-\cntt)^3 +\cntd\cntt +\cntt^ 3)$.
	
	Finally, for sufficiently large bubbles (close to the critical temperature $T_c$) the latter choice also ceases to be convenient. But we know that in this regime the solution is approximately homogeneous everywhere except close to the wall and, in particular, its profile is well approximated by a hyperbolic tangent. For this reason we choose the change
	\begin{equation} \label{eq:change3}
		\rho(x) = \rho_3(x) = \rho_0 + \wz \, \text{artanh}\parent{x + (x-1)\tanh\parent{\frac{\rho_0}{\wz}}}\,,
	\end{equation}
	designed to zoom in the relevant region of width $\wz$ around the radius $\rho_0$. The values of these parameters are calculated using the fits to a hyperbolic tangent in Eq.~\eqref{eq:radius_width1} from previously found solutions and extrapolating to the next, with\footnote{In hindsight, it would have probably been even better to choose $\wz = m\cdot\lw$, with $m$ some number between $2$ and $\rho_0/2$, in such a way that the coordinate does not zooms in only the wall but also a small region around it. Nevertheless, our choice worked well enough in our case.} $\rho_0 = \radius$ and $\wz = \lw$. There is however a subtlety in the coordinate choice defined by Eq.~\eqref{eq:change3}. Note that $\rho_3(x)$ is almost vertical at the origin, which makes it numerically challenging to impose Neumann boundary conditions there. Fortunately, when this choice becomes the convenient one, the solution at the origin is so close to the corresponding homogeneous stable state that it is possible to switch to Dirichlet boundary conditions there.
	
	In contrast, we did not need to perform any change of variables along the $z$ coordinate. Practically, we set $\zH = 1$, which implies that all quantities extracted from our numerical solution are expressed in units of $\zH$. Equivalently, this amounts to working with the coordinate $\mathbf{z} = z/\zH$.
	
	After all these definitions, we are left with a system of six second order PDEs defined in $(x,z)\in[0,1]\times [0,1]$. We used Chebyshev grids in both directions, with $n_x$ and $n_z$ points respectively, including the boundaries. This allowed us to solve the problem with grids of moderate size. However, the method is limited for two reasons. On the one hand, note the appearance of logarithms in our asymptotic expansions, Eq.~\eqref{eq:UV_expansion_inhomogeneous}, which spoil spectral convergence. Convergence is also limited by the development of the bubble wall, but this is cured by the different choices of coordinates introduced earlier. These two issues forced us to work with precision higher than \verb|MachinePrecision|, which is easily implemented in Mathematica. For most of the bubbles that we constructed, $n_z=50$ and $n_x  = 50$ or $80$, even though we checked robustness of the results by computing some of them again in a  $70\times 70$ grid. In most cases, we computed the reference metric with 80 digits of precision and solved the system of PDEs with 60 digits of precision, using a Newton-Raphson procedure, see Ref.~\cite{Krikun:2018ufr}. 
	\begin{figure}
		\centering
		\includegraphics[width=0.49\linewidth]{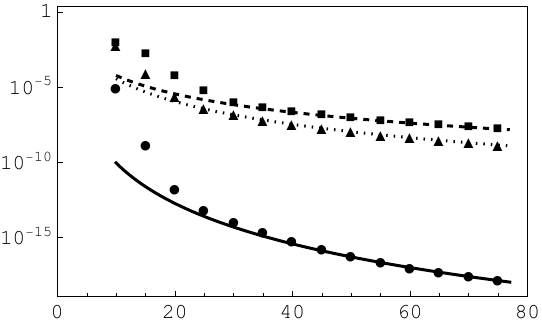}
		\put(-100,-13){$n_{\text{\tiny grid}}$}
		\put(-200,135){$\max|\mathcal{E}_i|$}\hfill \includegraphics[width=0.49\linewidth]
		{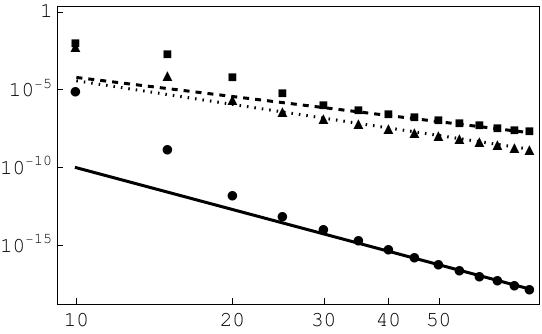}
		\put(-200,135){$\max|\mathcal{E}_i|$}
		\put(-100,-13){$n_{\text{\tiny grid}}$}
		\caption{\small Maximum absolute value of the different components of $\xi_A$, properly normalized, as a function of the number of grid points in every directions ($n_{\text{\tiny grid}} = n_x = n_z$). These plots show that spectral convergence is spoiled in our case, probably due to the presence of the logarithms near the boundary. The curves show fits for $\mathcal{E}_1$~(circles, solid curve), $\mathcal{E}_2$ (squares, dashed curve) and $\mathcal{E}_3$ (triangles, dotted curve) to $\max|\mathcal{E}_i|=K_i (n_{\text{\tiny grid}})^{a_i}$ using the last six last data points shown.}
		\label{fig:convergence}
	\end{figure}
	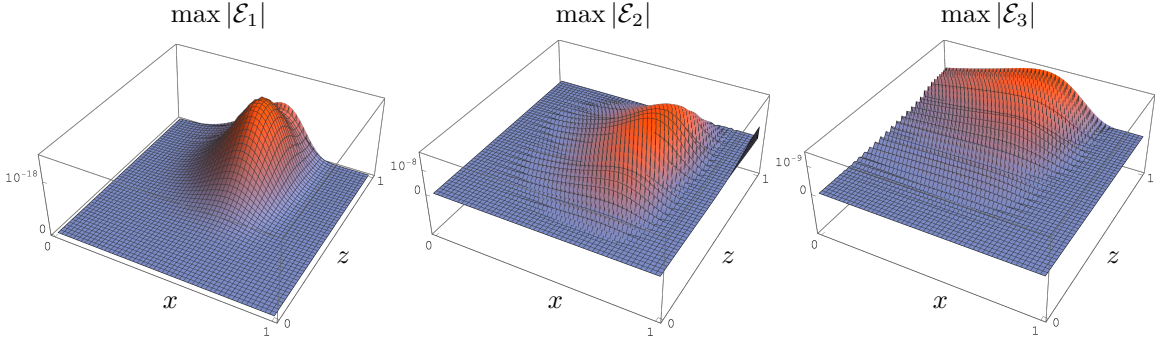
\begin{figure}
		\centering
		\includegraphics[width=0.33\linewidth]{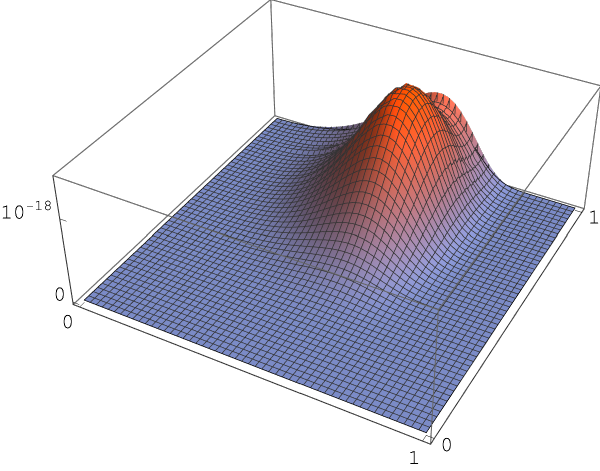}
		\put(-83,12){\small $x$}
		\put(-18,29){\small $z$}
		\put(-80,120){\small $\max|\mathcal{E}_1|$}\hfill \includegraphics[width=0.33\linewidth]{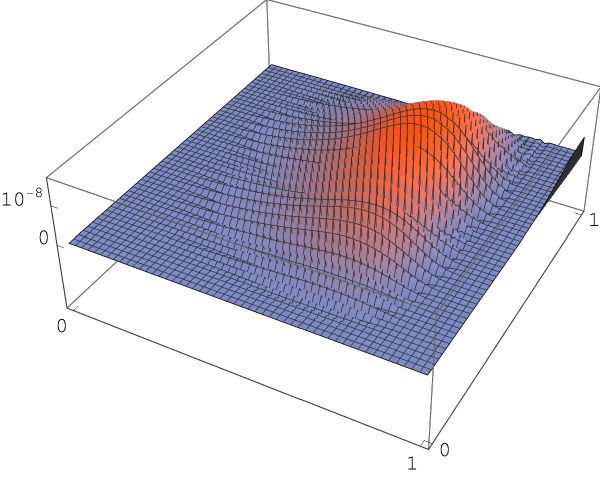}
		\put(-83,12){\small $x$}
		\put(-18,29){\small $z$}
		\put(-80,120){\small $\max|\mathcal{E}_2|$}\hfill\includegraphics[width=0.33\linewidth]{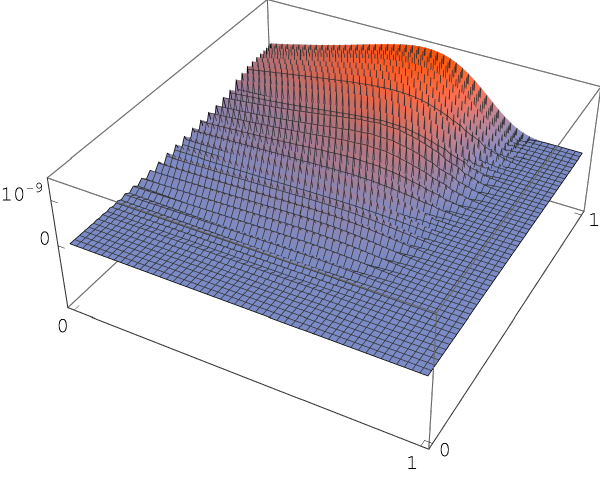}
		\put(-83,12){\small $x$}
		\put(-18,29){\small $z$}
		\put(-80,120){\small $\max|\mathcal{E}_3|$}
		\caption{\small Plot of the quantities defined on Eq.~\eqref{eq:app_errors}  for $n_{\text{\tiny grid}} = 75$.} \label{fig:errors_on_grid}
	\end{figure}
	
	In addition, the evaluation of the De Turk vector $\xi_A$ gives us a way to check the convergence of our implementation. Recall that the only non-vanishing components of $\xi_A$ as defined in Eq.~\eqref{eq:DeTurck_vector} were $\xi_\rho$ and $\xi_z$. Analyzing how these approach $\rho = 0$ and $\rho\to\infty$, we realized that it is convenient to define the quantities
	\begin{equation}\label{eq:app_errors}
		(\mathcal{E}_1,\mathcal{E}_2\, ,\mathcal{E}_3) = \parent{
			\quad
			\rho^2\xi_A\xi^A,\quad
			\rho\xi_\rho ,\quad
			(\zH+\rho)\xi_z 
			\quad
		}\,,
	\end{equation}
	where the additional factors of $\rho$ make sure that the corresponding $\mathcal{E}_i$ approach the origin and infinity \textit{off-shell} as
	\begin{equation}
		\mathcal{E}_i = \mathbf{e}_{i,0}(z) + O(\rho)\,,\qquad 
		\mathcal{E}_i = \mathbf{e}_{i,\infty}(z) + O(\rho^ {-1})\,.
	\end{equation}

	\section{Holographic Renormalization} \label{app:holoren}
	In this Appendix, we collect the formulas for the renormalized expectation values of the holographic energy-momentum tensor and the scalar operator, obtained after performing the holographic renormalization procedure (see Refs.~\cite{Bianchi:2001de,Bianchi:2001kw,Skenderis:2002wp}). We follow the structure of Appendix~B in Ref.~\cite{Ecker:2023uea}, with the caveat that $\rho|_{\text{\tiny there}} = \tilde z^2|_{\text{\tiny here}}$. The starting point is to write the metric in FG coordinates, as in Eq.~\eqref{eq:def:FG}, which we copy here for ease of reference,
	\begin{equation}\label{eq:FGansatz_appendix}
		\dd s^ 2 = \frac{L^ 2}{\zt^2}\parent{\dd \zt^ 2 + g_{\mu\nu}(\zt, x)\dd x^\mu\dd x^\nu}\,.\tag{see~\eqref{eq:def:FG}}
	\end{equation}
	Recall that the boundary is located at $\zt = 0$. For any gravity theory defined by Eq.~\eqref{eq:actionGH} and whose potential 
	behaves close to the origin as Eq.~\eqref{eq:expansion_potential}, the metric and the scalar field approach the boundary as 
	\begin{equation}
		\begin{aligned}
			\label{eq:UV_general_app}
			{g}_{\mu\nu}(\zt,x)&=\frac{L^2}{\zt^2} \parent{\gamma_{\mn}(x) \,+\,  \gamma_{(2)\mn}(x){\zt^2}\,+\, \parent{\gamma_{(4)\mn}(x) + h_{(4)\mn}(x)\log \zt }\zt^4  \,+\, O(\zt^6) }\,,
			\nonumber\\[\spacelist]
			\phi(\zt,x)&= \zt\parent{\phi_{(0)}(x)+ \parent{\phi_{(2)}(x)  + \psi_{(2)}(x) \log \zt }\zt^2  +O(\zt^4)}\,.   
		\end{aligned}
	\end{equation}
	Four of the coefficients we just wrote are not fixed by the asymptotic analysis of the equations. Two of these are the leading orders coefficients $\gamma_{\mn}(x)$ and $\phi_{(0)}(x)$, which correspond to the boundary metric and the source of the operator dual to~$\phi$, respectively. In our paper, $\gamma_{\mn}(x) = \eta_\mn$ was always fixed to flat Minkowski metric, while\footnote{With $J=0$ in Section~\ref{sec:holo_setup}.} $\phi_{(0)}(x) =  \fs + J(x)$. We will nevertheless keep the expressions in this Appendix general. While the asymptotic analysis imposes two conditions that the components of $\gamma_{(4)\mn}(x)$ need to satisfy, the remaining freedom and the coefficient $\phi_{(2)}(x)$ are only fixed after specifying regularity conditions in the bulk (in our case, the presence of a regular horizon). The rest of the terms in Eq.~\eqref{eq:UV_general_app} are given in terms of the two leading ones,
	\begin{equation}
		\label{eq:subleading_terms_app}
		\begin{aligned}
			\psi_{(2)}&=-\frac{1}{4} \left(\sqb\phi_{(0)}-\frac{1}{6}\phi_{(0)} \Rb\right)\,,\qquad  \gamma_{(2)\mn}=-\frac{1}{2}\left(\Rb_{\mn}-\frac{1}{6} \Rb \, \gamma_{\mn} \right)
			-\frac{\phi_{(0)}^2}{3} \gamma_{\mn}\,,\\[5mm]
			h_{(4)\mn}&=h_{(4)\mn}^{\mbox{\tiny{grav}}}-\frac{1}{12}\Rb_{\mn}\phi_{(0)}^2-\frac{1}{3}\nb_\mu\phi_{(0)}\nb_\nu\phi_{(0)}
			\, + \, 
			\frac{1}{12}\nb_\sigma\phi_{(0)}\nb^\sigma\phi_{(0)}\,\gamma_{\mn}+\frac{1}{6}\phi_{(0)}\nb_\mu\nb_\nu\phi_{(0)}
			\\\space &\qquad+ \, 
			\frac{1}{12}\phi_{(0)} \sqb\phi_{(0)}\, \gamma_{\mn}\,,\\\space
		\end{aligned}
	\end{equation}
	where the bars indicate that the corresponding geometrical quantity is computed with respect to the boundary metric $\gamma_\mn$, and
	\begin{equation}
		\begin{aligned}
			h_{(4)\mn}^{\mbox{\tiny{grav}}}&=\frac{1}{8}\Rb_{\mu\sigma\nu\tau}\Rb^{\sigma\tau}-\frac{1}{48}\nb_\mu\nb_\nu \Rb+\frac{1}{16}\sqb\Rb_{\mn}-\frac{1}{24}\Rb\Rb_{\mn} \\\space
			&\qquad +\left(\frac{1}{96}\Rb^2-\frac{1}{96}\sqb\Rb-\frac{1}{32}\Rb_{\sigma \tau}\Rb^{\sigma \tau}\right)\gamma_{\mn}\,.
		\end{aligned}
	\end{equation}
	
	Now, the renormalized action of our holographic model contains three pieces:
	\begin{equation}
		\label{eq:Sren_app}
		I_{\text{\tiny ren}} = I_g + I_{\text{\tiny GHY}}+ I_{\text{\tiny ct}} \,.
	\end{equation}
	The first term is the bulk action Eq.~\eqref{eq:actionGH}, while the second one is the usual Gibbons-Hawking-York (GHY) boundary term included to have a well posed variational problem,
	\begin{equation}
		\label{eq:GHY_app}
		I_{\text{\tiny GHY}} =  \frac{1}{\kappa_5^2}\int\dd^4x\sqrt{-h} K\,,
	\end{equation}
	with $h$ the determinant of $h_\mn$ the induced metric on a constant $\tilde z = \eUV$ four-dimensional slice near the boundary. At the end of the computation we take $\eUV\to 0$. Moreover, $K = h^\mn K_\mn $ is the trace of the extrinsic curvature $K_\mn$ of the four-dimensional slice. Finally, 
	\begin{equation}
		\label{eq:ct_app}
		\begin{aligned}
			I_{\text{\tiny ct}} &= \frac{2}{\kappa_5^2} \int \dd^4x\sqrt{-h}\Bigg[ - \frac{\Rh L}{8} -  \frac{3}{2L}  
			+\Bigg(\frac{L^3}{16}\parent{\Rh_\mn\Rh^\mn - \frac{1}{3}\Rh^2} + \frac{L}{2}\parent{\phi\tilde\square\phi-\frac{1}{6}\Rh\phi^2}\Bigg)\log\eUV \\[2mm]
			&\qquad\qquad\qquad\qquad + \alpha {\mathcal{A}}L^3+ \be\frac{\phi^4}{L} \Bigg]\,
		\end{aligned}
	\end{equation}
	is a counter-term  defined on the $\zt = \eUV$ hypersurface which makes $I_{\text{\tiny ren}}$ finite when the limit $\eUV\to 0$ is taken. In this expression the tildes indicate that the corresponding geometric quantities are computed using the metric $h_\mn$. Actually, the two last terms accompanied by the constants $\alpha$ and $\be$ are finite and stand for residual renomalization-scheme ambiguities of the model. The first one is proportional to
	the holographic conformal anomaly~\cite{Henningson:1998gx,Papadimitriou:2011qb} $\mathcal{A}=\mathcal{A}_g +\mathcal{A}_\phi$ consisting of a gravitational part due to the curved boundary geometry and a part due to scalar matter
	\begin{equation}\label{eq:anomaly_qpp}
		\mathcal{A}_g = \frac{1}{16}(\Rb^{\mn}\Rb_{\mn}-\frac{1}{3}\Rb^2)\,,\qquad
		\mathcal{A}_\phi=-\frac{1}{2}\left( \nb_\sigma\phi_{(0)}\nb^\sigma\phi_{(0)}+\frac{1}{6}\Rb\phi_{(0)}^2\right)\,.
	\end{equation}
	The second one plays the role of a vacuum energy.
	
	With all this information, we can give an expression for the expectation value of the boundary stress tensor,
	\begin{equation}\label{eq:EMT_app}
		\begin{aligned}
			&\langle  T_{\mn}^{\text{\tiny{QFT}}}\rangle=\frac{2}{\sqrt{-\gamma}}\frac{\delta I_{\rm ren}}{\delta \gamma^{\mn}}\\
			&= \frac{2L^3}{\kappa_5^2} 
			\Bigg[\gamma_{(4)\mn}+\frac{1}{8}\left(\mathrm{Tr}\gamma_{(2)}^2-(\mathrm{Tr}\gamma_{(2)})^2\right)\gamma_{\mn}
			-\frac{1}{2}\gamma_{(2)}^2+\frac{1}{4}\gamma_{(2)\mn}\mathrm{Tr}\gamma_{(2)}+\frac{1}{2}\nb_\mu\phi_{(0)}\nb_\nu\phi_{(0)}\\
			&+\left(\phi_{(0)}\phi_{(2)}-\frac{1}{2}\phi_{(0)}\psi_{(2)}-\frac{1}{4}\nb_\sigma\phi_{(0)}\nb^\sigma\phi_{(0)}\right)\gamma_{\mn}
			+\alpha\left(\mathcal{T}^{\gamma}_{\mn}+\mathcal{T}_{\mn}^\phi\right)+\left(\frac{1}{18}+\be \right)\phi_{(0)}^4\gamma_{\mn}\Bigg]\,,
		\end{aligned}
	\end{equation}
	with the anomalous contributions given by
	\begin{equation}
		\begin{aligned}
			\mathcal{T}^{g}_{\mn}&=2h_{(4)\mn}\,,\\\space
			\mathcal{T}_{\mn}^\phi&=-\frac{1}{6}\phi_{(0)}^2\bar{R}_{\mn}-\frac{2}{3}\nb_\mu\phi_{(0)}\nb_\nu\phi_{(0)}+\frac{1}{6}\nb_\sigma\phi_{(0)}\nb^\sigma\phi_{(0)}\gamma_{\mn} +\frac{1}{3}\phi_{(0)}\nb_\mu\nb_\nu\phi_{(0)}\\
			&-\frac{1}{3}\phi_{(0)}\sqb \phi_{(0)} \gamma_{\mn}+\frac{1}{12}\bar{R}\phi_{(0)}^2\gamma_{\mn} \,.
		\end{aligned}
	\end{equation}
	
	Similarly, the expectation value of the operator dual to $\phi$ is
	\begin{equation}\label{eq:VEV_app}
		\VEV=\frac{1}{\sqrt{-\gamma}}\frac{\delta S_{\rm hol}}{\delta \phi_{(0)}} = \frac{2L^3}{\kappa_5^2}
		\left[(1-4\alpha)\psi_{(2)}-2\phi_{(2)}-4\be\phi_{(0)}^3\right]\,,
	\end{equation}
	and the anomaly-corrected Ward identities
	\begin{equation}\label{eq:Ward_app}
		\nb^\mu\langle T_{\mn}^{\rm QFT}\rangle=-\VEV\nb_\nu \phi_{(0)}\,,\qquad
		\gamma^{\mn}\langle T_{\mn}^{\rm QFT}\rangle=- \phi_{(0)}\VEV+
		\frac{2L^3}{\kappa_5^2} \left(\mathcal{A}_{g}+\mathcal{A}_{\phi}\right)\,,
	\end{equation}
	are satisfied.

	\section{Free energy} \label{app:free energy}
	In this Appendix we show that  the on-shell action, which provides the value of the free energy, can be computed as in Eq.~\eqref{eq:free_energy1}. We start by showing different identities we will employ.
	
	\subsection{Useful expressions}
	The bubble solutions constructed in this work possess a hypersurface-orthogonal timelike Killing field $\zeta^A$. This means that,
	\begin{equation}\label{eq:basic_properties_zeta}
		\nabla_{(A}\zeta_{B)} = 0,\qquad \zeta_{[A}\nabla_B  \zeta_{C]}=0\,.
	\end{equation}
	In our coordinates ---Eqs. \eqref{eq:def:FG} and \eqref{eq:static_PDE}--- this vector field is just $\zeta = \partial_t$. In addition, our geometries have a (inhomogeneous) Killing horizon, generated by $\zeta^A$ on the hypersurface where its norm vanishes.
	
	Denoting the norm of the Killing vector as $\zeta^2 \equiv g_{AB}\zeta^A\zeta^B$, and defining $Z = \sqrt{-\zeta^2}$, we can find two useful expressions involving covariant derivatives of $\zeta^A$ along its integral curves,
	\begin{equation}\label{eq:expre_intermediate_3}
		\begin{aligned}
			\zeta^ B\nabla_B\zeta_A &= - \zeta^ B\nabla_A\zeta_B = -\frac{1}{2}\nabla_A(\zeta^2) = \frac{1}{2}\nabla_A(Z^2)=Z\nabla_A(Z)\,,\\[2mm]
			\zeta^B\nabla_B(Z)  & = \frac{1}{2Z}\zeta^B\nabla_B(Z^2) =-\frac{1}{2Z}\zeta^B\nabla_B(\zeta^C\zeta_C) = -\frac{1}{Z}\zeta^B\zeta^C\nabla_B\zeta_C = 0\,.
		\end{aligned}
	\end{equation}
	
	Some of the expressions that we will find later will involve the future directed unit vector field $u^A \equiv \zeta^A/\sqrt{-\zeta^2}= \zeta^A/Z$, whose acceleration vector $a^A \equiv u^B\nabla_Bu^A$ satisfies the important property 
	\begin{equation}\label{eq:expre_intermediate_2}
		a_A = u^B\nabla_Bu_A  =  u^B\nabla_B \parent{\frac{\zeta_A}{Z}} = \frac{1}{Z^2}\zeta^B\nabla_B\zeta_A - \frac{1}{Z^3}\zeta^B\zeta_A\nabla_B(Z) = \frac{1}{Z}\nabla_A(Z)\,.
	\end{equation}
	In the last step we used the two identities in Eq.~\eqref{eq:expre_intermediate_3}.
	
	So far we have only used that $\zeta^A$ is Killing. Contracting the hypersurface orthogonality condition $\zeta_{[A}\nabla_B\zeta_{C]}=0$ with $\zeta^C$ we obtain
	\begin{equation}
		\zeta^C \zeta_{A}\nabla_B  \zeta_{C} +
		\zeta^C\zeta_{B}\nabla_C  \zeta_{A}
		-Z^2\nabla_A  \zeta_{B} = 0\,.
	\end{equation}
	This, in turn, can be rewritten as
	\begin{equation}\label{eq:covariant_zeta_and_acceleration_anti}
		\nabla_A\zeta_B = a_A\zeta_B- a_B\zeta_A    
	\end{equation}
	making use of Eqs.~\eqref{eq:expre_intermediate_3} and~\eqref{eq:expre_intermediate_2}. Using this expression, together with Eq.~\eqref{eq:expre_intermediate_2}, we obtain the last property that we will use later,
	\begin{equation}\label{eq:covariant_zeta_and_acceleration_sym}
		\nabla_A\parent{\frac{\zeta_B}{\zeta^2}} = 
		-\nabla_A\parent{\frac{\zeta_B}{Z^2}} = -\frac{1}{Z^2}\nabla_A{\zeta_B} + \frac{2}{Z^3}\zeta_B\nabla_A Z 
		=\frac{1}{Z^2}\parent{a_A\zeta_B + a_B\zeta_A}\,.
	\end{equation}
	
	\subsection{Evaluation of the on-shell action}
	With the previous expressions at hand, let us evaluate Eq.~\eqref{eq:Sren_app} on our solutions. We focus our attention first on the bulk integral $I_g$, explicitly written in Eq.~\eqref{eq:actionGH}. After several steps we will be able to show that its integrand  can be written as a total derivative. Start by reordering Einstein's equations as
	\begin{equation}
		\begin{aligned}
			g_{MN} R = -2\parent{T_{MN}-R_{MN}} = -2\parent{2\partial_M\phi\partial_N\phi - g_{MN} \parent{(\partial_A\phi\partial^A\phi + 2V(\phi))}-R_{MN}}\,.\\[1mm]
		\end{aligned}
	\end{equation}
	As a consequence, projecting onto 
	$\zeta^A$ we obtain
	\begin{equation}
		R =2  \parent{(\partial_A\phi\partial^A\phi + 2V(\phi))}+\frac{2}{\zeta^2}\zeta^A\zeta^BR_{AB}\,,
	\end{equation}
	where we have used that $\zeta^A\partial_A\phi = \zeta^A\nabla_A\phi = 0$. Substituting this expression into the integrand of the bulk action $I_g$ in Eq.~\eqref{eq:actionGH}, we discover that it can expressed in terms of derivatives of the Killing vector,
	\begin{equation}\label{eq:rewrite_bulk_integrand}
		\frac{R}{4}-\frac12 \partial_A\phi\partial^A\phi
		-V(\phi) = \frac{1}{2\zeta^2} \zeta^A\zeta^B R_{AB} = - \frac{1}{2\zeta^2} \zeta_B\nabla_A\nabla^A \zeta^B\,,
	\end{equation}
	In the last identity we have used that, since $\zeta^A$ is Killing, $\nabla_A\nabla^A \zeta^B = -R^{B}_{\ A}\zeta^A$. 
	
	Now we would like to express Eq.~\eqref{eq:rewrite_bulk_integrand} as a total derivative, so that we can use Stokes' theorem to perform the integral. Integration by parts results into
	\begin{equation}\label{eq:integration_by_parts}
		- \frac{1}{2\zeta^2} \zeta_B\nabla_A\nabla^A \zeta^B = -\frac{1}{2}\left[ \nabla_A\parent{\frac{\zeta_B}{\zeta^2}\nabla^A\zeta^B} - \nabla_A \parent{\frac{\zeta_B}{\zeta^2}} \nabla^A {\zeta^B}\right]\,.
	\end{equation}
	Importantly, the second term inside the squared parenthesis is
	the contraction of the antisymmetric expression in Eq.~\eqref{eq:covariant_zeta_and_acceleration_anti} with the symmetric one from Eq.~\eqref{eq:covariant_zeta_and_acceleration_sym}. Therefore, that term vanishes in our case and the integrand in $I_g$ is indeed a total derivative. As a consequence we can use Stokes' theorem and write
	\begin{equation}\label{eq:IG_as_sum_boundaries}
		I_g 
		= \frac{2}{\kappa_5^2}\int\dd^5x\sqrt{-G}\,\nabla_A\parent{-\frac{\zeta_B}{2\zeta^2}\nabla^A\zeta^B} 
		= -\frac{1}{\kappa_5^2}\sum_{i}\int \parent{ \frac{\zeta_B}{\zeta^2} \nabla^A\zeta^B}\dd\Sigma^{(i)}_A \,.
	\end{equation}
	Here  the sum runs over the different boundaries, with 
	$\dd\Sigma^{(i)}_A$
	denoting each directed hypersurface volume element.
	For spacelike or timelike boundaries,
	\begin{equation}\label{eq:def.hdd.1}
		\dd\Sigma^{(i)}_A \equiv n_A^{(i)}\sqrt{|h^{(i)}|}\, \dd^4 y \equiv n_A^{(i)} \dd V^{(i)} \,,
	\end{equation}
	with $\dd V^{(i)}$ the induced volume element, $n_A^{(i)}$ the outward unit normal, and $h^{(i)}$  the determinant of the induced metric on the boundary, 
	\begin{equation}\label{eq:def.hdd.2}
		h_{MN}^{(i)} = g_{MN} - s\, n^{(i)}_M n^{(i)}_N\,,\qquad s \equiv n_An^A =\pm 1\,.
	\end{equation}
	In our solutions, there are only two boundaries  contributing to Eq.~\eqref{eq:IG_as_sum_boundaries}, the asymptotic AdS boundary and the horizon, $I_g = I_{\infty} + I_{\text{\tiny H}}$. Note that the horizon is a null hypersurface. As a consequence, we cannot use Eqs.~\eqref{eq:def.hdd.1} and \eqref{eq:def.hdd.2} directly. Rather, we will consider a family of timelike surfaces and take the limit  in which this family approaches the horizon. Also, to simplify the notation, we use the labels $i = \infty$ at the boundary of AdS and $i=\text{H}$ for the family of hypersurfaces close to the horizon. Moreover, we denote the corresponding outward unit normals as $n_A\equiv n_A^{(\infty)}$  and $m_A\equiv n_A^{(\text{\tiny H})}$.
	
	The boundary of AdS is the hypersurface at $z=0$ in our coordinates, Eq~\eqref{eq:static_PDE}. An outward normal is $\omega = - \dd z$, and consequently the outward unit-normal for this boundary reads $n_A = \omega_A/\sqrt{\omega_A\omega_B G^{AB}}$. Noting that it is perpendicular to the future directed unit vector field $u^A$ ---defined above Eq. \eqref{eq:expre_intermediate_2}--- it follows that 
	\begin{equation}\label{eq:expre_intermediate_1}
		u^B\nabla_B\parent{n_Au^A} = 0 = u^Au^B\nabla_Bn_A + n^Au^B\nabla_Bu_A = u^Au^B K_{AB} + n^Aa_A\,,
	\end{equation}
	since the extrinsic curvature is $K_{AB} =h^{C}_{\ A}h^{D}_{\ B} \nabla_Cn_D$. Hence, at the asymptotic boundary
	\begin{equation}\label{eq:UV_piece}
		\frac{1}{\zeta^2} \zeta_Bn_A\nabla^A\zeta^B = 
		-\frac{1}{Z^2} \zeta_B  n_A \nabla^A\zeta^B = 
		\frac{1}{Z} n_A \nabla^A(Z) = n_A a^A =  - u^Au^B K_{AB} =  K_t^ {\, t}\,,
	\end{equation}
	where we used Eqs.~\eqref{eq:expre_intermediate_3}, \eqref{eq:expre_intermediate_2} and \eqref{eq:expre_intermediate_1}. In the last equality we particularized the result to our choice of coordinates. We conclude that the integral at this boundary is
	\begin{equation}\label{eq:UV_piecebis}
		I_\infty = -\frac{1}{\kappa_5^ 2}\int\dd^4x\sqrt{-h}\  K_t^ {\, t}\,.
	\end{equation}
	This integral is divergent, but its divergences will be canceled by the GHY term and the counterterms, Eqs.~\eqref{eq:GHY_app} and \eqref{eq:ct_app}.
	
	On the other hand, the horizon lies at the  constant $z=\zH$ slice. The corresponding outward normal is $\lambda = + \dd z$, with outward unit-normal given by $m_A = \lambda_A/\sqrt{\lambda_B\lambda_C G^{BC}}$. The horizon contribution is then
	\begin{equation}\label{eq:int_hor_1}
		I_{\text{\tiny H}} = -\frac{1}{\kappa_5^2} \lim_{z\to\zH}\int { \frac{\zeta_B}{\zeta^2} \nabla^A\zeta^B}\dd\Sigma^{(\text{\tiny H})}_A  = -\frac{1}{\kappa_5^2} \lim_{z\to\zH}\int { \frac{\zeta_B}{\zeta^2} \nabla^A\zeta^B} m_A \sqrt{|h^{(\text{\tiny H})}|}\, \dd^4 y
	\end{equation}
	In coordinates for which $G_{MN}$ is block diagonal, such as the ones used in Eq.~\eqref{eq:static_PDE}, the square root of the determinant can be written as 
	$
	\sqrt{|h^{(\text{\tiny H})}|} = \sqrt{-G_{tt}}\cdot \sqrt{q^{(\text{\tiny H})}}\,,
	$
	and the integral at the horizon becomes
	\begin{equation}\label{eq:int_hor_2}
		I_{\text{\tiny H}} = \frac{1}{\kappa_5^2} \lim_{z\to\zH}\int \parent{ - \frac{\zeta_B}{\zeta^2} \nabla^A\zeta^B m_A \sqrt{-G_{tt}} }  \times  \sqrt{|q^{(\text{\tiny H})}|}\, \dd t\dd^3 y\,.
	\end{equation}
	When the limit is taken, the piece inside the parenthesis equals the surface gravity, $\kappa_{\text{\tiny H}}$. Secretly, this is because $\zeta^A$ is the horizon generator, and $m_A$ is perpendicular to it. In addition, the left over square root gives the horizon area density. Hence, 
	\begin{equation}\label{eq:int_hor_3}
		I_{\text{\tiny H}} = \frac{\kappa_{\text{\tiny H}}}{\kappa_5^2} \int \sqrt{|q^{(\text{\tiny H})}|}\, \dd t\dd^3 y = T S\int\dd t\,.
	\end{equation}
	
	Putting Eqs.~\eqref{eq:UV_piecebis} and~\eqref{eq:int_hor_3} together we obtain
	\begin{equation}\label{eq:result_free_energy_app}\begin{aligned}
			I_{\text{\tiny ren}}  &= \frac{1}{\kappa_5^2}\int\dd^4 x\sqrt{-h} \parent{{ - K_t^ {\ t}} + {K} +  \frac{2}{L}\parent{ - \frac{3}{2}  -  \frac{\phi^2}{2}  + 
					\be \phi ^4} } + TS \int\dd t  \,.
		\end{aligned}
	\end{equation}
	Now we perform a Wick rotation $t\mapsto - i\tau$ in Eq.~\eqref{eq:result_free_energy_app} to obtain the Euclidean action, $I_{\text{\tiny ren}}^{\text{\tiny (E)}}  = -iI_{\text{\tiny ren}}\big|_{t\, \mapsto\,  - i\tau}$. After integrating the compact Euclidean time coordinate $\tau\in(0,1/T)$,
	\begin{equation}\label{eq:result_free_energy_Euclidean_app}\begin{aligned}
			I_{\text{\tiny ren}}^{\text{\tiny (E)}} \cdot T  &= -\frac{1}{\kappa_5^2}\int\dd^3 x\sqrt{h_E} \parent{{- K_t^ {\ t}} + {K} +  \frac{2}{L}\parent{ - \frac{3}{2}  -  \frac{\phi^2}{2}  + 
					\be\phi ^4} } - T S  \,.
		\end{aligned}
	\end{equation}
	From this we obtain an expression for the free energy, which is precisely $ F= I_{\text{\tiny ren}}^{\text{\tiny (E)}} \cdot T$. Evaluating the integrand in Eq.~\eqref{eq:result_free_energy_Euclidean_app} using the UV expansions~\eqref{UVexp_alt} and \eqref{eq:UV_expansion_inhomogeneous}, we discover that the first term is just the energy, given in Eq.~\eqref{eq:expression_energy}. In conclusion, we obtain
	\begin{equation}
		F = E - TS\,.
	\end{equation}

	\section{Multiple pole contributions to the effective action}
	\label{app:effact}
	
	So far we have assumed there is a single pole in the correlator that is relevant for the effective action. In general, the correlator of a composite operator, such as the one appearing in the holographic model under consideration, does not exhibit a single isolated pole, but rather a nontrivial analytic structure involving multiple poles and branch-cut singularities, although the latter are usually absent in holographic models at nonzero temperature. In this case the correlator consists of a sum over poles 
	\begin{equation}
		G(\omega=0,q)=\sum_n \frac{R_n}{q^2-P_n} \,. 
	\end{equation}
	From \eqref{eq:effective_action_expansion}, the second derivative of the quantum effective action around a homogeneous equilibrium configuration is
	\begin{equation}\label{eq:d2gamma}
		\frac{\partial^2 \Gamma}{\partial \varphi^2}\Big|_{\varphi=\varphi_0}=
		\left( \sum_n \frac{R_n}{P_n} 
		\right)^{-1}
		=-\left( \frac{\partial^2 W}{\partial J^2}\right)^{-1}\Big|_{J=0}.
	\end{equation}
	On the other hand, a general solution to the equations of motion \eqref{eq:eomsGamma} is a superposition of modes corresponding to the poles of the correlator
	\begin{equation}
		\varphi(x)=\varphi_0+\sum_n \eta_n(x),\quad  \left(-\partial_i^2   -P_n^2\right)\eta_n(x)=0.
	\end{equation}
	We can obtain this same equations of motion from a quadratic effective action with a field for each mode. Taking $Z_n=-1/R_n$ and $ M_n^2= P_n^2/R_n$,
	\begin{equation}
		S_{\text{\tiny eff}}=\int d^4x\sum_n\left( \frac{1}{2}Z_n (\partial_i\eta_n)^2+\frac{1}{2}M_n^2 \eta_n^2+V_0\right).
	\end{equation}
	Beyond quadratic order, but to second order in derivatives, 
	\begin{equation}
		S_{\text{\tiny eff}}=\int d^4x \left(\sum_n \frac{1}{2}Z_n(\vec{\eta})(\partial_i\eta_n)^2+\Veff(\vec{\eta})\right),
	\end{equation}
	where we have grouped the fields in a vector $\vec{\eta}=(\eta_1,\eta_2,\cdots)$. Extrema of the potential $\Veff$ correspond to equilibrium states. By construction, $\vec{\eta}=0$ is an extremum, but other equilibrium states would lie away from the origin of field space. Expanding around a different equilibrium state one will find similar expressions, but the basis of modes that diagonalizes the potential will be different in general. 
	
	Even though this is a much more complicated situation than the one of a single pole, it is reasonable to assume that the profile of non-homogeneous solutions like the critical bubbles would follow a preferred direction in field space connecting two equilibrium points, presumably along the directions where the potential is less steep. We can estimate the steepness around the origin in each direction from the Hessian
	\begin{equation}\label{eq:Hessian_components}
		H_{nm}=\frac{\partial^2 V}{\partial\varphi_n\partial\varphi_m}=M_n^2\delta_{nm}=
		\frac{P_n}{R_n}\delta_{nm}.
	\end{equation}
	Comparing to \eqref{eq:d2gamma}, we also observe that the second derivative of the quantum effective action is determined by the trace of the inverse of the Hessian  
	\begin{equation}\label{eq:d2gammaHess}
		\frac{\partial^2 \Gamma}{\partial \varphi^2}\Big|_{\varphi=\varphi_0}=\frac{1}{{\rm Tr}( H^{-1})} \,. 
	\end{equation}
	The less steep direction correspond to the smallest eigenvalue of the Hessian. If one of the eigenvalues becomes parametrically small, it will dominate in \eqref{eq:d2gammaHess}, and the second derivative of the effective potential will coincide with the value predicted by a single independent field,
	\begin{equation}
		\eta_{\rm min}=\sum_n a_n \eta_n, \quad H \eta_{\rm min}=\lambda_{\rm min}\eta_{\rm min},\quad \frac{\partial^2 \Gamma}{\partial \varphi^2}\Big|_{\varphi=\varphi_0}\approx \lambda_{\rm min}\,.
	\end{equation}
	In some special cases the field can correspond to the contribution of a single pole, rather than a superposition. This actually happens at the spinodal points, where two extrema of the effective potential coalesce, $\partial_\field^2 \Veff(\varphi_\ho,T)$ vanishes, and one mode becomes gapless. In our analysis, we assume that single field dominance is the relevant scenario for bubble configurations.
	
	\subsection{The effect of contact terms when several poles are considered}
	
	In the model we examined, the presence of finite contact terms had an important influence in constructing the effective action. Let us end this Appendix by pointing out how the equations we just found are altered by the presence of these. The Green function is now
	\begin{equation}
		G(\omega=0,q)=\sum_n \frac{R_n}{q^2-P_n} +\sum_i c_i q^{2i},
	\end{equation}
	and Eq.~\eqref{eq:d2gamma} becomes
	\begin{equation}\label{eq:d2gammacontact}
		\frac{\partial^2 \Gamma}{\partial \varphi^2}\Big|_{\varphi=\varphi_0}=
		\left( \sum_n \frac{R_n}{P_n} - c_0
		\right)^{-1}
		=-\left( \frac{\partial^2 W}{\partial J^2}\right)^{-1}\Big|_{J=0}.
	\end{equation}
	Note that only the $c_0$ term contributes, as we are evaluating these expressions on homogeneous configurations. The components of the Hessian at the extrema \eqref{eq:Hessian_components} are not altered, and thus Eq.\eqref{eq:d2gammaHess} in presence of the contact terms becomes
	\begin{equation}\label{eq:d2gammaHesswithc0}
		\frac{\partial^2 \Gamma}{\partial \varphi^2}\Big|_{\varphi=\varphi_0}=\frac{1}{{\rm Tr}( H^{-1})-c_0} \,.
	\end{equation}
	Importantly, when the dominating pole becomes almost gapless, $c_0$ becomes negligible compared to the inverse Hessian and it can be ignored to first approximation.

	\section{Calculation of the two-point function} \label{app:greenfunc}
	In order to compute the two-point function we perturb the homogeneous solution, introducing fluctuations of the metric and the scalar.  It is convenient to work  we do in Eddington-Finkelstein coordinates as in Ref.~\cite{Jansen:2017oag},
	\begin{equation}
		\dd s^2 = -F(u)\dd t^2 + 2H(u) \dd u\dd t + S(u)^2 (\dd x_1^2+\dd x_2^2+\dd x_3^2)\,.
	\end{equation}
	This ansatz for the metric is related to Eq.~\eqref{eq:static} via
	\begin{equation}
		H(u)\dd u = \frac{L^2}{z^2}\, f(z)^{\frac{1}{2}}\dd z,\qquad F(u) = \frac{L^2}{z^2}f(z),\qquad S(u) = \frac{L}{z}g(z)^{\frac{1}{2}}\,.
	\end{equation}
	Generically, the fluctuations take the form
	\begin{equation}
		G_{\mu\nu}\mapsto G_{\mu\nu} + \epsilon \,h_{\mu\nu}(t,\vec x,u) + O(\epsilon^2),\qquad \phi \mapsto \phi + \epsilon\, \delta\phi(t,\vec x,u) + O(\epsilon^2)\,,
	\end{equation}
	but we will be interested in plane wave perturbations, and so we take the simplified ansatz
	\begin{equation}
		h_{\mu\nu}(t,x_1,u) = e^{-i(\omega t - q x_1)} h_{\mu\nu}(u)\,,\qquad 
		\delta\phi(t,x_1,u) = e^{-i(\omega t - q x_1)} \delta \phi(u)\,.
	\end{equation}
	
	\newcommand{\Zfl}{\mathsf{Z}}
	In our case, only scalar perturbations matter for the computation of the effective action. The two relevant gauge invariant combinations are\footnote{Note that, compared to \cite{Jansen:2017oag}, $\Zfl_2|_{\text{there}} = q^2 \Zfl_2|_{\text{here}}$.}
	\begin{equation}
		\Zfl_2 =  h_{tt} + 2\frac{\omega}{q} h_{tx} + \frac{\omega^2}{q^2} h_{xx} + \left( -\frac{\omega^2}{q^2} + \frac{F'}{2SS'} \right) h\,,\qquad
		\Zfl_\phi = \delta \phi - \frac{\phi'}{2SS'} h \,,
	\end{equation}
	with $h = (h_{22} + h_{33})/2$ and primes here indicating derivatives with respect to $u$. Since we are interested in constructing critical bubbles, which are static configurations, from now on we set $\omega=0$. The solutions then approach the boundary of spacetime as
	\begin{equation}\label{eq:expansion_UV}
		\begin{aligned}
			\Zfl_\phi &=a_1 z+\parent{ a_3 + \parent{\frac{a_1 q^2}{2}-b_{0}\frac{\fs}{12 L^2}}\log z}z^3+O\left(z^4\right)\,.\\
			\Zfl_2 &=\frac{1}{z^2}\parent{b_0-\frac{1}{12} b_0 \left(3 q^2+4\fs^2\right)z^2+\parent{b_4-\frac{b_0 q^2}{48}  \left(3 q^2+4\fs^2\right) \log (z) }z^4+O\left(z^6\right)}\,.
		\end{aligned}
	\end{equation}
	We emphasize  that $\fs$ is the coupling, given by the leading order in the background scalar field, see Eq.~\eqref{UVexp_alt}.
	
	Since the kinetic term appears as a gradient expansion of the fluctuations, we focus on the limit of small momenta, $q\to 0$. Therefore we expand
	\begin{equation}\label{eq:expansion_q}
		\Zfl_2(z) = \Zfl_2^{(0)}(z) + q^2 \Zfl_2^{(2)}(z) + O(q^3)\,,\qquad \Zfl_\phi(z) = \Zfl_\phi^{(0)}(z) + q^2 \Zfl_\phi^{(2)}(z) + O(q^3)\,.
	\end{equation}
	Later, we will see that it is precisely in taking this small $q$ limit that some subtleties arise.
	Let us for now relate Eqs.~\eqref{eq:expansion_UV} and \eqref{eq:expansion_q} by substituting the latter into the equations for the fluctuations. By doing so, we find a system of four second order ODEs for $\Zfl_i^{(0)}(z)$ and $\Zfl_i^{(2)}(z)$ which does not depend on $q$. Close to the boundary the solution for each function in the momentum expansion becomes
	\begin{equation}\begin{aligned}
			\Zfl_\phi^{(0)}(z) &= a_{1,0} z+a_{3,0} z^3+O\left(z^4\right)\,,\\
			\Zfl_2^{(0)}(z) &= \frac{1}{z^2}\parent{b_{0,0}-  \frac{b_{0,0}}{3}  \fs^2z^2+b_{4,0} z^4+O\left(z^5\right)}\,,\\[\spacelist]
			\Zfl_\phi^{(2)}(z) &= a_{1,2}z +\parent{a_{3,2} + \parent{\frac{a_{1,0}}{2}-\frac{b_{0,0} \fs }{12 L^2}}\log(z)}z^3+O\left(z^4\right)\,,\\
			\Zfl_2^{(2)}(z) &= \frac{1}{z^2}\parent{b_{0,2}-\frac{1}{12} z^2 \left(4 \fs ^2 b_{0,2}+3 b_{0,0}\right)+\frac{1}{12} z^4 \left(12 b_{4,2}-b_{0,0}
				\fs ^2 \log (z)\right)+O\left(z^5\right)}\,,\\[\spacelist]
		\end{aligned}
	\end{equation}
	Note that $a_i = a_{i,0} + a_{i,2}q^2 + O(q^4)$, with $i=1,\,3$; and $b_i = b_{i,0} + b_{i,2}q^2 + O(q^4)$ with $i=0,\,4$.
	
	Because we are not varying the boundary metric, we set $b_0=b_{0,0}=b_{0,2}=0$. The two point function is now nicely given in terms of the formula obtained from holographic renormalization, Eq.~\eqref{eq:VEV_app}, which in this case becomes
	\begin{equation}\label{eq:variations}
		\begin{aligned}
			\delta\VEV &= \VEV - \VEV|_{\epsilon=0}  = -\frac{\epsilon}{12}   e^{i q x_1} \left(8 \left(a_1 \fs ^2 (18 \be +1)+3 a_3\right)+3 (4 \alpha -1) a_1 q^2\right)\,,\\[\spacelist]
			\delta J &= \phi_{(0)}(x) - \fs = a_1 \epsilon  e^{i q x_1}\,.
		\end{aligned}
	\end{equation}
	Note the explicit appearance of $\alpha$, which boils down to the fact that the source for the leading order of the scalar $\phi_{(0)}(x)$ is no longer constant and, consequently, $\psi_{(2)}\neq 0$, see Eq.~\eqref{eq:subleading_terms_app}.
	
	From the variations in Eq.~\eqref{eq:variations} we can easily compute the two point function at the homogeneous equilibrium states, i.e.~the Green's function:
	\begin{equation}\label{eq:two-point-function}
		\begin{aligned}
			\widetilde G_0 &=\langle \mathcal{O}\mathcal{O}\rangle = \frac{\delta\VEV}{\delta J}\Big|_{J=0} = -\frac{8 \left(a_1 \left(18 \be +1\right) \fs ^2+3 a_3\right)+3 (4 \alpha -1) a_1
				q^2}{12 a_1}\\
			&= \left( -\frac{2 \, a_{3,0}}{a_{1,0}} - \frac{2}{3} \fs^2 \left(1 + 18 \, \be \right) \right) + \left( \frac{1}{4} + \frac{2 \, a_{1,2} \, a_{3,0}}{a_{1,0}^2} - \frac{2 \, a_{3,2}}{a_{1,0}} - \alpha \right) q^2 + O (q^4)\,.
		\end{aligned}
	\end{equation}
	The inverse of this two point function is
	\begin{equation}\label{eq:gamma2}
		\begin{aligned} 
			\widetilde G_0^{-1} &= -\frac{3 a_{1,0}}{2 \left( 3 a_{3,0} + a_{1,0} \fs^2 (1 + 18 \be \right)}
			+ \frac{9 \left( -8 a_{1,2} a_{3,0} + 8 a_{1,0} a_{3,2} + a_{1,0}^2 ( 4 \alpha-1) \right) q^2}
			{16 \left( 3 a_{3,0} + a_{1,0} \fs^2 + 18 a_{1,0} \fs^2 \be \right)^2}
			+ O(q^4)\,,
		\end{aligned}
	\end{equation}
	which give us an expression for the coefficients appearing in Eq.~\eqref{eq:small_expansion_Green_function},
	\begin{equation}\label{eq:Zalpha_dependent}
		\begin{aligned}
			A_0(\be) &= -\frac{3 a_{1,0}}{2 \left( 3 a_{3,0} + a_{1,0} \fs^2 (1 + 18 \be \right)}\\
			B_0(\alpha,\be)  &= - \frac{9 \left( -8 a_{1,2} a_{3,0} + 8 a_{1,0} a_{3,2} + a_{1,0}^2 ( 4 \alpha-1) \right)}
			{16 \left( 3 a_{3,0} + a_{1,0} \fs^2 + 18 a_{1,0} \fs^2 \be\right)^2}\,.    
		\end{aligned}
	\end{equation}
	If we identified $B_0(\alpha,\be)$ with (minus) the kinetic term, we would have a dependence not only on $\be$ but also on $\alpha$. Instead, we compute the kinetic term from the pole of the correlator $P_0$, as in Eq.~\eqref{eq:polevalues}. Using that the second derivative of the potential is given by $-A_0(\be)$,  we conclude that the kinetic term is related to the pole as
	\begin{equation}
		\label{eq:kinetic term_final}
		Z(\field) = -\frac{1}{R_0} = - \frac{3 a_{1,0}}{2 \left( 3 a_{3,0} + a_{1,0} \fs^2 (1 + 18 \be) \right)} \times P_0^{-1}\,.\\
	\end{equation} 
	
	\bibliographystyle{JHEP}
	\bibliography{main.bib}
	\vspace{1cm}
\end{document}